\documentclass[useAMS,usenatbib]{mn2e}

\usepackage{graphicx}
\usepackage{amssymb}

\newcommand{\ms}{\scriptscriptstyle}
\newcommand{\tnm}{\textnormal}

\title[Detection Thresholds for Linear Polarization]{
Analytic Detection Thresholds for Measurements of Linearly
Polarized Intensity Using Rotation Measure Synthesis
}
\author[Hales et al.]{
C.~A. Hales$^{1,2}$\thanks{E-mail: c.hales@physics.usyd.edu.au},
B.~M. Gaensler$^{1,3}$,
R.~P. Norris$^{2}$ and
E. Middelberg$^{4}$\\
$^{1}$Sydney Institute for Astronomy, School of Physics, The University of Sydney, NSW 2006, Australia\\
$^{2}$CSIRO Astronomy \& Space Science, PO Box 76, Epping, NSW 1710, Australia\\
$^{3}$Australian Laureate Fellow\\
$^{4}$Astronomisches Institut, Ruhr-Universit\"{a}t Bochum, Universit\"{a}tsstr. 150, 44801 Bochum, Germany}
\begin{document}


\pagerange{\pageref{firstpage}--\pageref{lastpage}} \pubyear{2012}

\maketitle

\label{firstpage}

\begin{abstract}
A fully analytic statistical formalism does not yet exist to describe radio-wavelength
measurements of linearly polarized intensity that are produced using rotation
measure synthesis. In this work we extend the analytic formalism for standard
linear polarization, namely that describing measurements of the quadrature sum of
Stokes $Q$ and $U$ intensities, to the rotation measure synthesis environment.
We derive the probability density function and expectation value for Faraday-space
polarization measurements for both the case where true underlying polarized
emission is present within unresolved Faraday components, and for the limiting
case where no such emission is present. We then derive relationships to quantify
the statistical significance of linear polarization measurements in terms of
standard Gaussian statistics. The formalism developed in this work will be
useful for setting signal-to-noise ratio detection thresholds
for measurements of linear polarization, for the analysis of polarized sources
potentially exhibiting multiple Faraday components, and for the development
of polarization debiasing schemes.

\end{abstract}


\section{Introduction}\label{sec:intro}
Radio-wavelength observations of linearly polarized synchrotron emission enable
studies of ionized gas and magnetic fields in, and along the lines-of-sight to,
energetic astrophysical environments. Faraday rotation measure (RM) synthesis
\citep{1966MNRAS.133...67B,2005A&A...441.1217B} is a
technique for Fourier transforming observational polarimetric data to produce a
complex Faraday dispersion spectrum. The magnitude of this spectrum, which
we denote by $|\mathcal{F}(\alpha,\delta,\phi)|$, encapsulates the intensity
of linearly polarized emission exhibited at different Faraday depths\footnote{Faraday
depth is not a physical depth, but rather the depth of Faraday rotating
magnetised plasma between a source of polarized emission and the
telescope; see equation~(3) from \citet{2005A&A...441.1217B}.},~$\phi$,
along a single physical line of sight with sky coordinate~$(\alpha,\delta)$.

Statistics describing measurements of linearly polarized intensity derived
from $|\mathcal{F}(\alpha,\delta,\phi)|$ have been investigated empirically by
\citet{2011arXiv1106.5362G} and analytically by
\citet{2012ApJ...750..139M}. However, a fully analytic
description is yet to be presented. Such statistics are required to enable
detailed quantitative analysis of polarimetric data from existent radio
facilities such as the Australia Telescope Compact Array Broadband Backend
\citep{2011MNRAS.416..832W}, the Westerbork Synthesis Radio Telescope,
the Giant Metrewave Radio Telescope, and the Expanded Very Large
Array \citep{2011ApJ...739L...1P}, and from future surveys such as POSSUM
\citep{2010AAS...21547013G} with the ASKAP observatory
\citep{2008ExA....22..151J,2009IEEEP..97.1507D}, GALFACTS
\citep{2010ASPC..438..402T} with the Arecibo observatory, and the
Magnetism Key Science
Project\footnote{http://www.mpifr-bonn.mpg.de/staff/rbeck/MKSP/mksp.html}
with the Low Frequency Array.

The statistics exhibited by Faraday-space measurements of linear
polarization are qualitatively similar, yet in general quantitatively
different, to those of standard linear polarization. Intensity measurements
of the latter, denoted\footnote{
The term $L$ is commonly used to differentiate linear polarization from the
more general elliptical polarization $P\equiv\sqrt{Q^2+U^2+V^2}$, which
includes Stokes $V$. While we do not use $P$ in this work, we follow the $L$
notation to ensure consistency with future deep surveys in which many sources
exhibiting both $L$ and $P$ emission are likely to be detected.}
by $L$, are obtained for a given line of sight by taking the quadrature sum of
measured Stokes $Q$ and $U$ intensities, namely
\begin{equation}\label{eqn:L}
	L(\alpha,\delta) \equiv
	\sqrt{[Q(\alpha,\delta)]^2+[U(\alpha,\delta)]^2} \, .
\end{equation}
The statistics of $L$ and discussion of detection thresholds are well
documented \citep{rice,1985A&A...142..100S,vla161,2006PASP..118.1340V}.
In contrast, intensity measurements of Faraday-space linear polarization
must be obtained by first devising a method to extract some, or all, of the
polarized emission that may be present over a range of Faraday
depths ($\phi$) within the Faraday dispersion spectrum ($\mathcal{F}$)
for a given line of sight. The specifics of this extraction process will dictate
the resulting polarization measurement statistics. In this work we focus on
measurements produced by extracting the peak\footnote{In practice, the peak
should be fitted to minimise pixel discretisation errors; for example, see
discussion of 3-point parabolic fits by \citet{halesB}.} intensity from a cleaned
\citep{2009A&A...503..409H} Faraday dispersion spectrum, denoted by
$\mathcal{F}^{cln}$ (note that this is not the clean component spectrum,
but rather the cleaned spectrum which contains convolved clean components
plus residuals), namely
\begin{equation}\label{eqn:LrmDEF}
	L_{RM}(\alpha,\delta) \equiv \tnm{max}
	\left(|\mathcal{F}^{cln}(\alpha,\delta,\phi)|\right) \,,
\end{equation}
where we use the term $L_{\ms RM}$ to differentiate these measurements from
those of standard $L$. Measurements of $L_{\ms RM}$, as defined in
equation~(\ref{eqn:LrmDEF}), are suitable for the analysis of data consisting
of unresolved\footnote{
Just as the peak surface brightness of an unresolved source in a two-dimensional
(2D) image (measured in Jy~beam$^{-1}$) is equal in magnitude to its integrated
surface brightness (or flux density; measured in Jy), so too is the peak polarized
intensity of an unresolved component in Faraday space \citep[measured in
Jy~beam$^{-1}$~RMSF$^{-1}$, where RMSF is the rotation measure spread
function, or the unit of resolution in Faraday space;][]{2005A&A...441.1217B}
equal in magnitude to its Faraday-integrated polarized intensity (measured in
Jy~beam$^{-1}$), modulo any statistical or measurement-induced biases.}
polarized emission in Faraday space; such
conditions are often encountered observationally due to the limited bandwidth
capabilities of many present-day telescopes \citep[e.g.][]{2009A&A...503..409H}
or the underlying physics of target sources (e.g. pulsars). Like $L$, $L_{\ms RM}$
is positive semi-definite ($\ge0$) and exhibits non-Gaussian statistics.

In this paper we seek to relate the statistical significance of
measurements of $L_{\ms RM}$ with those of $L$ and of standard
Gaussian statistics for general observational setups, in
order to facilitate detailed quantitative analysis. To meet this aim we
analytically, rather than empirically, derive the probability density
function (PDF) and expectation value for measurements of $L_{\ms RM}$
for the general case where true underlying polarized emission is present,
and for the limiting case where no such emission is present.
For comparison, we note that \citet{2011arXiv1106.5362G}
have presented an empirical investigation of detection thresholds and the PDF
for $L_{\ms RM}$ for a specific observational setup; we seek to formally
generalise these results here. Additionally, through
comparison with simulated data, \citet{2012ApJ...750..139M} identified
a missing correction factor in their analytic PDF that we derive here.

We begin in \S~\ref{sec:standard} by reviewing the existing analytic
statistical description of $L$, and by deriving a relationship that
equates the significance of detections in $L$ with those of Gaussian
statistics. In \S~\ref{sec:rmstats} we extend these analytic results
to $L_{\ms RM}$, noting two key experiment-specific
parameters that dictate its observed statistical properties. In
\S~\ref{sec:ex} we demonstrate use of our derived significance relationships
for $L$ and $L_{\ms RM}$ through worked examples, and discuss how individual
lines of sight exhibiting multiple unresolved Faraday-space components may
be treated. In \S~\ref{sec:psp} we use our analytic results to illustrate
how cross-sectional profiles for astronomical sources with 2D elliptical
Gaussian morphologies are affected when observed in images exhibiting
polarization measurement statistics, namely, where each pixel in an
image of $L$ or $L_{\ms RM}$ is formed using
equation~(\ref{eqn:L}) or (\ref{eqn:LrmDEF}), respectively.
Using these profiles we briefly outline challenges
for robust source extraction in polarization images.
In \S~\ref{sec:ngQU} we address the point recently raised by
\citet{2011arXiv1106.5362G} that non-Gaussianities in images of
Stokes $Q$ and $U$ will complicate the calculation of robust
significance relationships. We present our conclusions and comment
on future work regarding polarization bias in \S~\ref{sec:conc}.

For notational convenience throughout this work we will drop the
explicit $(\alpha,\delta)$ notation (cf. equations above), but note
that the statistics we discuss refer to the distribution of intensities
that an individual line of sight (or pixel) may exhibit. As pointed out
in \S~\ref{sec:standard}, this is not necessarily the same as
discussing the statistics of a sample of measurements from different lines
of sight, or pixel intensities within some spatial region of an image.

\section{Standard Linear Polarization}\label{sec:standard}

The magnitude of observed standard linear polarization, $L$, is given by
equation~(\ref{eqn:L}). For true underlying Stokes intensities $Q_{\ms 0}$
and $U_{\ms 0}$ in the presence of Gaussian measurement errors
$\sigma_{\ms Q}$ and $\sigma_{\ms U}$, respectively, the observed
Stokes intensities $Q$ and $U$ as used in equation~(\ref{eqn:L}) are
given by
\begin{eqnarray}
	Q &=& Q_{\ms 0} \pm \sigma_{\ms Q} \, , \label{EQNAStatsQ}\\
	U &=& U_{\ms 0} \pm \sigma_{\ms U} \, . \label{EQNAStatsU}
\end{eqnarray}
The true underlying and unbiased linearly polarized signal is given by
\begin{equation}\label{eqn:L0}
	L_{\ms 0} = \sqrt{Q_{\ms 0}^{\ms 2}+U_{\ms 0}^{\ms 2}} \, .
\end{equation}

Measurements of Stokes $Q$ and $U$ may be obtained at radio wavelengths
using either individual spectral channel observations, resulting in
individual measurements of $L$ for each observed channel,
or band-averaged (e.g. using multi-frequency synthesis) observations,
resulting in a single measurement of $L$ for the entire band. Analytically,
there is no need to differentiate between these two approaches; the
analytic statistical descriptions of $L$ for the individual channel and
band-averaged approaches are identical; both can be described by
equations~(\ref{EQNAStatsQ}) and (\ref{EQNAStatsU}) for a sample of
measurements with given $L_{\ms 0}$. For completeness, we note that
the technique of RM-synthesis is preferable to the band-averaged $L$
approach because the latter is more prone to bandwidth depolarization,
in which rotations of spectral $Q$ and $U$ measurements through the
complex plane can cause their band-averaged values to become diminished.
Separately, we note that discussion regarding the statistics of stacked
measurements of $L$ is beyond the scope of this work; for example, as a
result of summing multiple $L$ measurements from individual spectral
channels over which Faraday rotation may be occurring.

The PDF for $L$ is given by \citet{rice} as
\begin{equation}\label{eqn:Rice}
	f\!\left(L|L_{\ms 0}\right)
	= \frac{L}{\sigma_{\ms Q,U}^{\ms 2}}
	\exp\!\left(-\frac{L^{\ms 2}+L_{\ms 0}^{\ms 2}}{2\sigma_{\ms Q,U}^{\ms 2}}\right)
	I_{\ms 0}\!\left(\frac{L L_{\ms 0}}{\sigma_{\ms Q,U}^{\ms 2}}\right) ,
\end{equation}
where $\sigma_{\ms Q,U}$ is a noise term explained below, $L\ge0$, $I_{\ms k}(x)$ is a
modified Bessel function of the first kind with order $k$ and argument $x$,
and it is assumed that the true polarized intensity $L_{\ms 0}$ ($\ge0$) is known. In
equation~(\ref{eqn:Rice}) and in future use, we simplify notation by assuming
that measurement error is implicitly specified in all priors; for example,
we imply $f(L|L_{\ms 0}) \equiv f(L|L_{\ms 0},\sigma_{\ms Q,U})$.
Equation~(\ref{eqn:Rice}) is known as the Ricean distribution. Formally, it
is only valid for $\sigma_{\ms Q,U}=\sigma_{\ms Q}=\sigma_{\ms U}$. We
assume this to be the case here, and discuss issues regarding
$\sigma_{\ms Q}\neq\sigma_{\ms U}$ in Appendix~\ref{sec:appA}. The Ricean
distribution is displayed for several values of the ratio
$L_{\ms 0}/\sigma_{\ms Q,U}$ in the top panel of Fig.~\ref{fig:statsALL}.

The cumulative distribution function (CDF) for $L$ is obtained by integrating
equation~(\ref{eqn:Rice}), giving
\begin{equation}\label{eqn:RiceCDF}
	F\!\left(L|L_{\ms 0}\right) = 1 - \mathcal{Q}_{\ms 1}\!\left( \frac{L_{\ms 0}}{\sigma_{\ms Q,U}},
	\frac{L}{\sigma_{\ms Q,U}} \right),
\end{equation}
where $\mathcal{Q}_{\ms 1}(\alpha,\beta)$ is the Marcum Q-function \citep{marcum}.
$\mathcal{Q}_{\ms 1}(\alpha,\beta)$ is defined by
\begin{equation}\label{eqn:Marcum}
	\mathcal{Q}_{\ms 1}\!\left(\alpha,\beta\right) = \int_{\beta}^{\infty} x
		\exp\!\left( -\frac{x^{\ms 2}+\alpha^{\ms 2}}{2} \right)
		I_{\ms 0}\!\left( \alpha x \right) \tnm{d}x \,,
\end{equation}
and may be efficiently calculated using the algorithm presented by \citet{simon}.

\citet{rice} gives both the expectation value (E) and variance (Var) of
equation~(\ref{eqn:Rice}) as
\begin{eqnarray}
	\tnm{E}(L|L_{\ms 0}) &=& \sqrt{\frac{\pi \sigma_{\ms Q,U}^{\ms 2}}{2}}\, 
		_{\ms 1}F_{\ms 1}\!\left( -\frac{1}{2};\,1;\,
		-\frac{L_{\ms 0}^{\ms 2}}{2\sigma_{\ms Q,U}^{\ms 2}} \right),\, \tnm{and}\label{eqn:RiceM}\\
	\tnm{Var}(L|L_{\ms 0}) &=& L_{\ms 0}^{\ms 2} + 2\sigma_{\ms Q,U}^{\ms 2} - 
		\left[\tnm{E}(L|L_{\ms 0})\right]^{\ms 2}\!, \label{eqn:RiceV}
\end{eqnarray}
where $_{\ms 1}F_{\ms 1}$ is a confluent hypergeometric function. In the absence of
input signal (i.e. $L_{\ms 0}=0$), the Ricean distribution limits to the \citet{rayleigh}
distribution given by
\begin{equation}\label{eqn:Ray}
	f\!\left(L|L_{\ms 0}=0\right)
	= \frac{L}{\sigma_{\ms Q,U}^{\ms 2}}
	\exp\!\left(-\frac{L^{\ms 2}}{2\sigma_{\ms Q,U}^{\ms 2}}\right) ,
\end{equation}
where $L\ge0$. The expectation value and variance of the Rayleigh distribution is given by
\begin{eqnarray}
	\tnm{E}(L|L_{\ms 0}=0) &=& \sqrt{\frac{\pi\sigma_{\ms Q,U}^{\ms 2}}{2}}
		\,\,,\, \tnm{and}\label{eqn:RayM}\\
	\tnm{Var}(L|L_{\ms 0}=0) &=& (4-\pi)\frac{\sigma_{\ms Q,U}^{\ms 2}}{2}
		\, . \label{eqn:RayV}
\end{eqnarray}
The CDF for the Rayleigh distribution is obtained by integrating
equation~(\ref{eqn:Ray}), giving
\begin{equation}\label{eqn:RayCDF}
	F\!\left(L|L_{\ms 0}=0\right) = 1 - 
	\exp\!\left(-\frac{L^{\ms 2}}{2\sigma_{\ms Q,U}^{\ms 2}}\right).
\end{equation}

Unlike a Gaussian distribution, the Ricean distribution is signal
(i.e. $L_{\ms 0}$) dependent; the Ricean PDF changes shape depending on
the magnitude of the underlying input signal (in comparison, the shape of
a Gaussian distribution is not influenced by terms in its PDF that
relate to its true or observed mean). The Ricean
distribution is positively skewed (right-skewed) and leptokurtic (positive
excess kurtosis) for weak input signal, while for stronger input signal the
distribution becomes Gaussian about mean $L_{\ms 0}$ with standard deviation
$\sigma_{\ms Q,U}$. It is this signal dependence that prevents one from
assuming a uniform variance (i.e. from assuming that $\tnm{Var}(L|L_{\ms 0})$
from equation~(\ref{eqn:RiceV}) is uniform) for different lines of sight
that have equal $\sigma_{\ms Q,U}$ (for example, a sample of spatial
pixels with equal $\sigma_{\ms Q,U}$ in an image of $L$, where the intensity
for each pixel is calculated using equation~(\ref{eqn:L})). The signal
dependence also complicates estimation of $L_{\ms 0}$ given a measurement
of $L$ \citep[e.g.][]{1985A&A...142..100S,vla161,2006PASP..118.1340V}.

We note that while the PDF for $L$ is non-Gaussian, the {\it noise} in a
Ricean distribution (represented by $\sigma_{\ms Q,U}$) is Gaussian in
character, reflecting the nature of measurement uncertainty in Stokes $Q$
and $U$ (such that $\sigma_{\ms Q,U}$ characterises the manner in which
random errors are propagated into measurements of $L$). The term {\it Ricean
noise} therefore has the potential to be misleading, as it may incorrectly
suggest that the Ricean distribution exhibits those properties usually
associated with regular Gaussian noise, such as signal-independence. If
the term Ricean noise is used, then the prefix {\it Ricean} should be
interpreted in the same way that, for example, {\it shot} noise, which
is governed by Poissonian statistics and is thus signal-dependent,
differentiates itself from standard Gaussian noise.

\subsection{Detection Significance}\label{sec:standardDS}

To quantify the significance of a measurement of $L$ in terms of a well-recognised
statistic, we relate its probability for Type~I (false positive) error to that of an
equivalent measurement of intensity in Gaussian noise. We use the term ``equivalent''
to indicate that the same noise term $\sigma_{\ms Q,U}$ from $L$ is used as the
standard deviation for the Gaussian distribution. We define the signal-to-noise ratio
(SNR) of a measurement of $L$ as $L/\sigma_{\ms Q,U}$. This definition makes use
of the observable quantities $L$ and $\sigma_{\ms Q,U}$; we do not relate $L$ to
equation~(\ref{eqn:RiceV}), which includes the unobservable and $L_{\ms 0}$-dependent
term $\tnm{E}(L|L_{\ms 0})$. We generically denote a measurement of intensity in
Gaussian noise by $G$ (e.g. a measurement of Stokes $Q$ intensity), and define
the SNR of our equivalent Gaussian measurement as $G/\sigma_{\ms Q,U}$. By
equating the CDF for a Rayleigh distribution [equation~(\ref{eqn:RayCDF})] with
the standard confidence interval for a Gaussian
(i.e. $\tnm{erf}[|G|/(\sqrt{2}\sigma_{\ms Q,U})]$, not its CDF), and by selecting
the magnitude of $G$ as the appropriate equivalent measure to compare with $L$,
we quantify the Gaussian equivalent significance, denoted by $G^{\ms ES}$,
for a measurement of $L$ as
\begin{equation}\label{eqn:LtoG}
	|G^{\ms ES}|/\sigma_{\ms Q,U} \equiv \sqrt{2}\,\tnm{erf}^{\ms -1}\!\left\{ 1 - \exp\!
		\left[\, -\frac{1}{2}\left(\frac{L}{\sigma_{\ms Q,U}}\right)^{\ms \!2}\,
		\right] \right\} \,,
\end{equation}
or, conversely, the linear polarization equivalent significance, denoted
by $L^{\ms ES}$, for a measurement of $G$ as
\begin{equation}\label{eqn:GtoL}
	L^{\ms ES}/\sigma_{\ms Q,U} \equiv \sqrt{-2\ln\!\left[1- \tnm{erf}
		\left(\frac{1}{\sqrt{2}}\,\frac{|G|}{\sigma_{\ms Q,U}}\right)\right]} \,,
\end{equation}
where $\tnm{erf}$ and $\tnm{erf}^{\ms -1}$ are the error function and
its inverse, respectively.

The Gaussian equivalent significance relationships above may be used to set
SNR cutoffs for polarization surveys, designed to meet the same statistical
criteria as standard $G/\sigma$ SNR cutoffs in surveys with Gaussian noise.
Examples illustrating use of these equations are presented in \S~\ref{sec:ex}.
Equation~(\ref{eqn:GtoL}), and thus implicitly equation~(\ref{eqn:LtoG}),
is displayed in Fig.~\ref{fig:detsig}.

\section{Faraday-Space Linear Polarization}\label{sec:rmstats}

In this section we derive customised statistics to describe measurements
of $L_{\ms RM}$ obtained using equation~(\ref{eqn:LrmDEF}). We begin by
discussing two experiment-specific parameters that will
be needed for this derivation: $M$, which characterises
an effective sample size in Faraday space, and $\sigma_{\ms RM}$, which
characterises the noise in $L_{\ms RM}$.

RM-synthesis can be thought of as a technique to evaluate $\mathcal{F}(\phi)$
over a range of trial Faraday depths spanning $\pm\phi_{\ms max}$, which
may be set by [equations (35) and (63) from \citealt{2005A&A...441.1217B}]
\begin{equation}\label{eqn:BB1}
	\phi_{\ms max} \approx \frac{\sqrt{3}}{\min \left[ \delta(\lambda^{\ms 2}_{\ms i}) \right]} \,,
\end{equation}
where $\delta(\lambda^{\ms 2}_{\ms i})$ are spectral channel widths in
wavelength-squared ($\lambda^{\ms 2}$) space for each $i$'th observed channel; the
minimum $\delta(\lambda^{\ms 2}_{\ms i})$ characterises the maximum Faraday
depth $\phi_{\ms max}$ at which polarized emission can be detected. The
effective resolution in Faraday space is
[equation (61) from \citealt{2005A&A...441.1217B}]
\begin{equation}\label{eqn:BB2}
	\psi \approx \frac{2\sqrt{3}}{\Delta(\lambda^{\ms 2})} \,,
\end{equation}
which is set by the observed wavelength-squared range
$\Delta(\lambda^{\ms 2})=\lambda_{\ms max}^{\ms 2}-\lambda_{\ms min}^{\ms 2}$.
\citet{2005A&A...441.1217B} note that equations~(\ref{eqn:BB1}) and (\ref{eqn:BB2})
assume a top hat weight function that is unity between $\lambda_{\ms min}$ and
$\lambda_{\ms max}$ and zero elsewhere. In general this will not be the case [cf.
equations~(\ref{eqn:RMnoise}) and (\ref{eqn:RMweights}) presented shortly],
requiring both $\phi_{\ms max}$ and $\psi$ to be determined empirically.
For example, $\psi$ may be fit with a Gaussian; this is analogous to
fitting a Gaussian to an experiment-specific point spread function in aperture
synthesis imaging. Combining equations~(\ref{eqn:BB1}) and (\ref{eqn:BB2}),
we find that $|\mathcal{F}^{cln}(\phi)|$ is effectively comprised of
\begin{equation}\label{eqn:Mdef}
	M \equiv \frac{2\phi_{\ms max}}{\psi}
\end{equation}
independent samples, as was recognised by both
\citet{2011arXiv1106.5362G} and \citet{2012ApJ...750..139M}.
In other words, no more than $M$ statistically independent measurements of
linearly polarized intensity may be extracted from $|\mathcal{F}^{cln}(\phi)|$,
assuming $\mathcal{F}$ is sampled with at least one trial $\phi$ per
resolution element $\psi$. However, this description of $M$
independent samples is only formally correct for ideally deconvolved signals.
It is not appropriate for describing noise, which will consist of $M$ independent
samples that have been permanently correlated with one another, due to the filtering
nature of the discrete Fourier transform underlying the RM-synthesis technique. The
presence of such correlations must be addressed to ensure a complete statistical
description of $L_{\ms RM}$. Further below, we decribe how the noise term
$\sigma_{\ms RM}$ may be defined so as to account for such correlations, enabling
the notion of $M$ independent samples to be effectively maintained. We note that the
statistics describing measurements of $|\mathcal{F}^{cln}(\phi_t)|$ for some fixed
trial Faraday depth $\phi_t$ (using the subscript $t$ momentarily for clarity), are
given by those of $L$ from \S~\ref{sec:standard}. This is because for each trial
$\phi_t$, RM-synthesis essentially unwraps the observed spectral $Q$ and $U$ data
in the complex plane so that their band-averaged values may be used to compute
$\mathcal{F}(\phi_t)$. Denoting the number of observed spectral channels
by $T$, we note that the form of equation~(\ref{eqn:BB1}) ensures that
$M>T$ for $T>1$; only for the trivial case $T=1$ does $M=T$.

In this work we assume that $\mathcal{F}$ consists of unresolved
(Faraday-thin) components. We also assume that $\mathcal{F}^{cln}$ has
been cleaned \citep[e.g. with {\tt RM-CLEAN};][]{2009A&A...503..409H}
in an idealised manner (which may not be met in practise for RM spread
functions exhibiting strong sidelobes) to prevent components in the spectrum
from being contaminated by sidelobes from other components. Therefore,
$L_{\ms RM}$ as defined in equation~(\ref{eqn:LrmDEF}) can
be characterised as the maximum of $M$ independent samples within a cleaned Faraday
dispersion spectrum, each of which exhibits the statistics of $L$ discussed
in \S~\ref{sec:standard}, and each of which may or may not contain any true
underlying signal $L_{\ms 0}$. For completeness, we note that the sequential
processing techniques of RM-synthesis and deconvolution require signal sparsity
in $\phi$-space \citep[cf. aperture synthesis imaging and the image sparsity
requirement of the {\tt CLEAN} technique;][]{1999ASPC..180..151C}. For
this work, we therefore require that the majority of $M$ independent samples
in $\mathcal{F}$ are signal-free (i.e. with $L_{\ms 0}=0$). We do not
consider the analysis of non-sparse Faraday dispersion spectra.

We denote the noise term for $L_{\ms RM}$ by $\sigma_{\ms RM}$. We define this
term as [note equation (38) from \citealt{2005A&A...441.1217B}]
\begin{equation}\label{eqn:RMnoise}
	\sigma_{\ms RM} = \left[\frac{1}{\eta}\,
	\frac{\sum_{\ms i=1}^{\ms T}w_{\ms i}^{\ms 2} \, \sigma_{\ms Q,U,i}^{\ms 2}}
	{\left(\sum_{\ms i=1}^{\ms T}w_{\ms i}\right)^{\ms 2}}\right]^{\!\ms \frac{1}{2}} ,
\end{equation}
where $\sigma_{\ms Q,U,i}$ is the noise in the $i$'th channel, 
$\eta$ is a correction factor described shortly, and
$w_{\ms i}$ are weighting factors for the observational data in each $i$'th
channel. For example, the channel weights may be chosen using least squares,
\begin{equation}\label{eqn:RMweights}
	w_{\ms i} = \frac{1}{\sigma_{\ms Q,U,i}^{\ms 2}}\,,
\end{equation}
noting as in \S~\ref{sec:standard} that the analysis in this section is only
formally valid for $\sigma_{\ms Q,U,i}=\sigma_{\ms Q,i}=\sigma_{\ms U,i}$ in
each $i$'th channel; see Appendix~\ref{sec:appA} for discussion regarding
$\sigma_{\ms Q,i}\neq\sigma_{\ms U,i}$. The factor $\eta$ in
equation~(\ref{eqn:RMnoise}) is required to account for correlations between
samples of $|\mathcal{F}^{cln}(\phi)|$ at different depths $\phi$. For
clarity, we note that if $L_{\ms RM}$ were defined by
$|\mathcal{F}^{cln}(\phi_t)|$ for fixed $\phi_t$, then $\eta=1$ would be
appropriate because issues regarding selection of the maximum of $>1$
correlated samples would be inapplicable.
Moving on, we assume an experimental setup where the total number of
trial $\phi$ samples across $\mathcal{F}$ is given by
\begin{equation}\label{eqn:kappa}
	\kappa = \frac{2\phi_{\ms max}}{\delta(\phi)}+1 \,,
\end{equation}
where each sample is spaced apart by $\delta(\phi)$; i.e. a sampling
rate of $\kappa/M$ per $\psi$. The autocorrelation function is given
by the magnitude of the RMSF for positive Faraday depths, which we denote by
$|R_h|\equiv|R[h\,\delta(\phi)\ge0]|$ with integer index $h=0,\ldots,\kappa-1$,
assuming an RMSF with span $\pm2\phi_{\ms max}$. Given positive correlation
between samples, as will always be the case given $|R_h|$, estimates of
$\sigma_{\ms RM}$ obtained using equation~(\ref{eqn:RMnoise}) with $\eta=1$
will always underestimate the true value. A correction for this bias
is given by \citeauthor{anderson} [\citeyear{anderson}; see equation~(51)
in chapter~8, adjusted to represent sample variance] as
\begin{equation}\label{eqn:G}
	\eta = 1 - \frac{2}{\kappa-1}\sum_{h=1}^{\kappa-1}
	\left(1-\frac{h}{\kappa}\right)|R_h| \,.
\end{equation}

Using the details above, we now derive the PDF for $L_{\ms RM}$ in the context of
order statistics \citep[e.g.][]{david}, first assuming $L_{\ms 0}=0$ in all $M$ samples,
then extending to the scenario where 1 of $M$ samples in the Faraday dispersion
spectrum contains an underlying signal $L_{\ms 0}>0$ (i.e. a single Faraday-thin
component). We will not extend this derivation to the more general scenario in
which each independent sample in the Faraday dispersion spectrum may have its
own independent value of $L_{\ms 0}\ge0$ (i.e. multiple Faraday-thin components),
though in principle the relevant PDF for this situation could be derived using
elements from the derivations below. However, we do discuss detection thresholds
for this scenario in \S~\ref{sec:exMUFC}. Additionally, we will not attempt to
derive PDFs for fully general scenarios in which resolved polarized emission in
Faraday space (i.e. Faraday-thick components) may be present\footnote{A systematic
positive bias in Faraday space, similar to that referred to in the image plane
as peak bias by \citet{halesB}, will need to be accounted for when measuring the
peak polarized intensity for resolved (Faraday-thick) sources.}.

For a sample of $N$ independent and identically-distributed variates
$X_{\ms 1},X_{\ms 2},\ldots,X_{\ms N}$ ordered such that
$X_{\ms (1)}<X_{\ms (2)}<\ldots<X_{\ms (N)}$ (using notation $X_{\ms (j)}$ for
ordered variates and $X_{\ms j}$ for unordered variates), then $X_{\ms (k)}$
is known as the $k$'th order statistic and $X_{\ms (N)}=\tnm{max}(X_{\ms j})$.
If $X$ has PDF $f(X)$ and CDF $F(X)$, then \citet{david} give the PDF for
$X_{\ms (k)}$ as
\begin{equation}
	f\!\left(X_{\ms (k)}\right) = \frac{N!}{\left(k-1\right)!\left(N-k\right)!}
		\frac{\left\{F\left[X_{\ms (k)}\right]\right\}^{\ms k-1}}
		{\left\{1-F\left[X_{\ms (k)}\right]\right\}^{\ms k-N}}
		f\left[X_{\ms (k)} \right]
		\,.\label{eqn:statsOrder}
\end{equation}

\begin{figure}
 \includegraphics[trim = 0mm 0mm 0mm 0mm, clip, angle=-90, width=84mm]{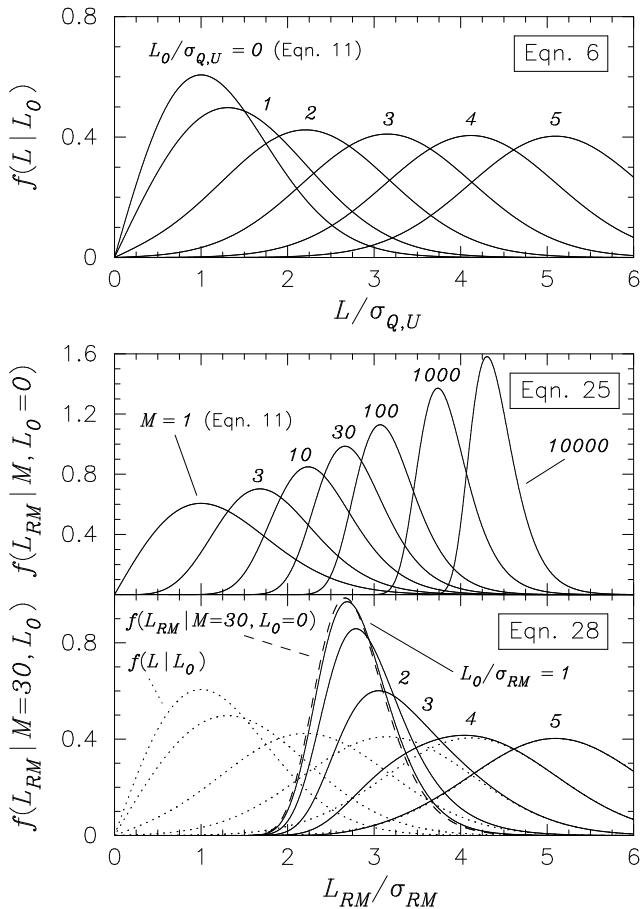}
 \caption{
	  {\it Top:} The Ricean distribution, equation~(\ref{eqn:Rice}), displayed
	             for several values of true polarization SNR $L_{\ms 0}/\sigma_{\ms Q,U}$.
		     When $L_{\ms 0}=0$ the Ricean distribution limits to the Rayleigh
		     distribution, equation~(\ref{eqn:Ray}).
	  {\it Middle:} The distribution for $L_{\ms RM}$
	  		with $L_{\ms 0}=0$, equation~(\ref{eqn:OrderRay}), displayed
	 		for several values of $M$, the effective number of independent samples in
			$|\mathcal{F}^{cln}(\phi)|$ as defined by equation~(\ref{eqn:Mdef}). This panel
			illustrates how different values of $M$ affect the mean value of
			$L_{\ms RM}$ for lines of sight free from polarized emission
			($L_{\ms 0}=0$). When $M=1$, equation~(\ref{eqn:OrderRay})
			limits to the Rayleigh distribution, equation~(\ref{eqn:Ray}).
	  {\it Bottom:} The distribution for $L_{\ms RM}$, equation~(\ref{eqn:Lrm}),
	  		displayed for the $M=30$ case (as suitable for a
			1.4~GHz observation with 200~MHz bandwidth
			split into 24 spectral channels) for several SNRs
			$L_{\ms 0}/\sigma_{\ms RM}$. As $L_{\ms 0}/\sigma_{\ms RM}$
			increases, equation~(\ref{eqn:Lrm}) limits to the Ricean
			distribution (dotted curves, replicated from top panel).
			However, unlike a Ricean distribution, as
			$L_{\ms 0}/\sigma_{\ms RM}\rightarrow0$, equation~(\ref{eqn:Lrm})
			limits to the signal-free distribution from
			equation~(\ref{eqn:OrderRay}) (dashed curve, replicated
			from middle panel).}
 \label{fig:statsALL}
\end{figure}

Assuming absence of an underlying input signal ($L_{\ms 0}=0$), the PDF
for $L_{\ms RM}$ is derived by substituting equations~(\ref{eqn:Ray}) and
(\ref{eqn:RayCDF}) into equation~(\ref{eqn:statsOrder}) with $N=M$ and $k=M$, giving
\begin{eqnarray}
	f\!\left(L_{\ms RM}|M,L_{\ms 0}=0\right) &=& M\,
		\frac{L_{\ms RM}}{\sigma_{\ms RM}^{\ms 2}}\,
		\exp\!\left(-\frac{L_{\ms RM}^{\ms 2}}{2\sigma_{\ms RM}^{\ms 2}}\right)
		\times\nonumber\\
		&&\left[ 1 - \exp\!\left(-\frac{L_{\ms RM}^{\ms 2}}{2\sigma_{\ms RM}^{\ms 2}}\right)
		\right]^{\ms \!M-1} \,,\label{eqn:OrderRay}
\end{eqnarray}
where $L_{\ms RM}\ge0$.
Equation~(\ref{eqn:OrderRay}) is displayed for several values of $M$ in the
middle panel of Fig.~\ref{fig:statsALL}; as $M$ increases, for example as a
result of increasing the spectral resolution in an experiment, so do the
resulting measured intensities. The expectation
value\footnote{We note that a derivation of the expectation value for the
signal-free case is attempted by \citet{2009A&A...503..409H}, where they
equate the $(M-1)/M$ quantile of the CDF with the expected value of the
largest order statistic. A better approximation is $M/(M+1)$; see
equation~(4.5.1) from \citet{david}. We present the exact solution in
equation~(\ref{eqn:OrderRayM}).} for equation~(\ref{eqn:OrderRay}) is
obtained using integration by parts, a Taylor expansion, and term-wise
integration, giving
\begin{eqnarray}
	\tnm{E}(L_{\ms RM}|M,L_{\ms 0}=0) &=& \int_{\ms 0}^{\ms \infty} L_{\ms RM}\,
		f\!\left(L_{\ms RM}|L_{\ms 0}=0\right) \, \tnm{d}L_{\ms RM} \nonumber\\
	&=& \sqrt{\frac{\pi\sigma_{\ms Q,U}^{\ms 2}}{2}}  \sum_{\ms S=1}^{\ms M} \big[
		S^{\ms -\frac{1}{2}} \left( -1 \right)^{\ms S-1} \times \nonumber \\
	&&	\frac{M!}{\left( M-S \right)!\,S!}\, \Big] . \label{eqn:OrderRayM}
\end{eqnarray}
Equation~(\ref{eqn:OrderRayM}) limits to equation~(\ref{eqn:RayM}) when $M=1$.
Equation~(\ref{eqn:OrderRayM}) represents the mean value of $L_{\ms RM}$
that will be observed for a line of sight containing no polarized emission
($L_{\ms 0}=0$). Asymptotically, this mean value grows as
$\approx\sqrt{\ln{M}}$ \citep{david}.

We now extend our derivation to the scenario where $M-1$ independent
signal-free samples are drawn from equation~(\ref{eqn:Ray}) and 1 sample
with arbitrary $L_{\ms 0}>0$ is drawn from equation~(\ref{eqn:Rice}),
such that $L_{\ms RM}$ represents the observed maximum of these $M$
samples. The distribution for the maximum intensity value of $M-1$
signal-free samples, which we denote $L_{\ms M-1}$, is derived in the
same manner as equation~(\ref{eqn:OrderRay}), but with $N=k=M-1$.
Following \citet{david}, the CDF for $L_{\ms RM}=\max(L_{\ms M-1},L)$
is then given by
\begin{equation}\label{eqn:LrmCDF}
	F\!\left(L_{\ms RM}|M,L_{\ms 0}\right) = F\!\left(L_{\ms M-1}|M,L_{\ms 0}=0\right) 
		F\!\left(L | L_{\ms 0}\right)\,.
\end{equation}
The PDF for $L_{\ms RM}$ is therefore
\begin{eqnarray}
	f\!\left(L_{\ms RM}|M,L_{\ms 0}\right)
	&=& 	\frac{\tnm{d}}{\tnm{d}L_{\ms RM}}
		F\!\left(L_{\ms RM}|M,L_{\ms 0}\right)\nonumber\\
	&=& 	f\!\left(L_{\ms M-1}|M,L_{\ms 0}=0\right) F\!\left(L | L_{\ms 0}\right) +\nonumber\\
	&&	F\!\left(L_{\ms M-1}|M,L_{\ms 0}=0\right) f\!\left(L | L_{\ms 0}\right)\nonumber\\
	&=& 	\frac{L_{\ms RM}}{\sigma_{\ms RM}^{\ms 2}}
		\exp\!\left(-\frac{L_{\ms RM}^{\ms 2}}{2\sigma_{\ms RM}^{\ms 2}}\right)
		\times \nonumber \\
	&&	\left[ 1 - \exp\!\left(-\frac{L_{\ms RM}^{\ms 2}}{2\sigma_{\ms RM}^{\ms 2}}\right)
		\right]^{\ms \!M-1}  \times \nonumber\\
	&\,&	\Bigg\{ \!\left( M-1 \right) \left[ 1 -
		\exp\!\left(-\frac{L_{\ms RM}^{\ms 2}}{2\sigma_{\ms RM}^{\ms 2}}\right)
		\right]^{\ms \!-1} \times \nonumber \\
	&&	\left[ 1 - \mathcal{Q}_{\ms 1}\!\left( \frac{L_{\ms 0}}{\sigma_{\ms RM}},
		\frac{L_{\ms RM}}{\sigma_{\ms RM}} \right) \right] +\nonumber\\
	&\,&	\exp\!\left(-\frac{L_{\ms 0}^{\ms 2}}{2\sigma_{\ms RM}^{\ms 2}}\right)
		I_{\ms 0}\!\left(\frac{L_{\ms RM} L_{\ms 0}}{\sigma_{\ms RM}^{\ms 2}}\right) 
		\!\Bigg\} \,, \label{eqn:Lrm}
\end{eqnarray}
where $L_{\ms RM}\ge0$. Equation~(\ref{eqn:Lrm}) is displayed for several values of
$L_{\ms 0}/\sigma_{\ms RM}$ for an $M=30$ observing setup in the bottom panel of
Fig.~\ref{fig:statsALL}; $M=30$ has been selected for illustrative
simplicity, as suitable for a 1.4~GHz observation with 200~MHz bandwidth split into
24 spectral channels.
The expectation value for equation~(\ref{eqn:Lrm}),
\begin{equation}\label{eqn:LrmM}
	\tnm{E}(L_{\ms RM}|L_{\ms 0}) = \int_{\ms 0}^{\ms \infty} L_{\ms RM}\,
		f\!\left(L_{\ms RM}|L_{\ms 0}\right) \tnm{d}L_{\ms RM} \, ,
\end{equation}
does not appear to have an analytic solution; it may be evaluated numerically. As with the
Ricean distribution, the distribution for $L_{\ms RM}$ is signal-dependent, positively skewed,
and leptokurtic. When the magnitude of $L_{\ms 0}$ is comparable to the noise $\sigma_{\ms RM}$,
the distribution for $L_{\ms RM}$ will approach the signal-free case from
equation~(\ref{eqn:OrderRay}). For larger $L_{\ms 0}/\sigma_{\ms RM}$, the distribution
will approach the Ricean distribution from equation~(\ref{eqn:Rice}).

The results presented in this section provide a theoretical explanation for the
empirical curves presented in Fig.~4 of \citet{2011arXiv1106.5362G} for their
specific experimental setup. Parameterised by $M$, the equations above enable
statistical characteristics of $L_{\ms RM}$ to be quantified for a range of
experimental setups.

Furthermore, our results provide an explanation for the
discrepancy between the simulated and theoretical PDFs presented
in the lower panel of Fig.~6 from \citet{2012ApJ...750..139M},
in which the effects of correlation were not considered. To demonstrate,
we evaluated equation~(\ref{eqn:G}) for an RMSF representing an experimental
setup similar to that described by \citet{2012ApJ...750..139M}, with
$24\times8$~MHz channels between 1296 and 1480~MHz, and Faraday space sampling
given by $\delta(\phi)=5$~rad~m$^{-2}$ with $\phi_{\ms max}=4000$~rad~m$^{-2}$.
The result was $\sqrt{\eta}=0.935$. Thus \citet{2012ApJ...750..139M}
overestimated their SNRs by $\sim7\%$, consistent with their
observed discrepancy\footnote{Separately, we note that
\citet{2012ApJ...750..139M} defined noise per channel (i.e.
$\sigma_{\ms Q,U,i}$) as the quadrature sum of $\sigma_{\ms Q,i}$ and
$\sigma_{\ms U,i}$, so that $\sigma_{\ms Q,U,i}=\sqrt{2}\sigma_{\ms Q,i}$ for
the case $\sigma_{\ms Q,i}=\sigma_{\ms U,i}$. As mentioned above and in
\S~\ref{sec:standard}, a more appropriate definition is
$\sigma_{\ms Q,U,i}=\sigma_{\ms Q,i}=\sigma_{\ms U,i}$.}.
(Note that the diminished peak density in their simulated PDF is accounted
for by the Jacobian; their green curve is not normalised.)

\subsection{Detection Significance}\label{sec:rmstatsDS}

\begin{figure}
 \includegraphics[trim = 0mm 0mm 0mm 0mm, clip, angle=-90, width=84mm]{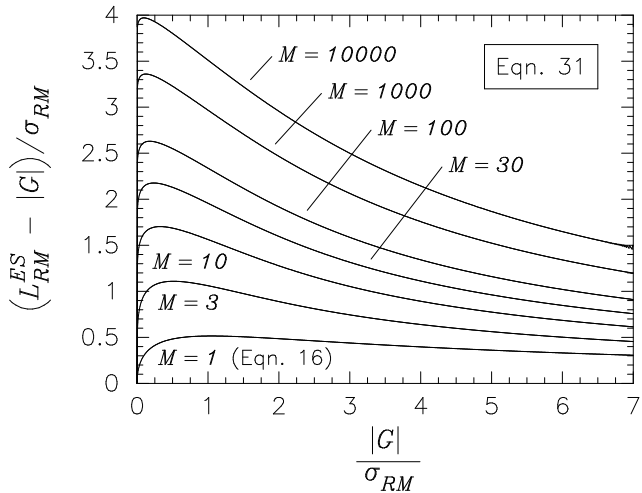}
 \caption{
	Detection thresholds for $L_{\ms RM}$ that exhibit equivalent
	Type~I (false positive) error rates to those of standard Gaussian
	detections, $G$. The curves trace equation~(\ref{eqn:GtoORay})
	for several values of $M$, limiting to equation~(\ref{eqn:GtoL})
	for $M=1$.
	}
 \label{fig:detsig}
\end{figure}

Following \S~\ref{sec:standardDS}, we quantify the Gaussian equivalent
significance, denoted by $G^{\ms ES}_{\ms RM}$, for a measurement of $L_{\ms RM}$
by equating the CDF for $L_{\ms RM}$ (i.e. $F(L_{\ms RM}|M,L_{\ms 0}=0)$,
obtained by integrating equation~(\ref{eqn:OrderRay})) with the standard
confidence interval for a Gaussian, giving
\begin{equation}\label{eqn:ORaytoG}
	|G^{\ms ES}_{\ms RM}|/\sigma_{\ms RM} \equiv \sqrt{2}\,\tnm{erf}^{\ms -1}\!\left( \left\{ 1 - \exp\!
		\left[\, -\frac{1}{2}\left(\frac{L_{\ms RM}}{\sigma_{\ms RM}}\right)^{\ms \!2}\,
		\right] \right\}^{\ms \!M} \right) .
\end{equation}
Alternatively, equation~(\ref{eqn:ORaytoG}) may be rearranged to
quantify the linear polarization equivalent significance, denoted by
$L_{\ms RM}^{\ms ES}$, for a measurement of $G$, giving
\begin{equation}\label{eqn:GtoORay}
	L_{\ms RM}^{\ms ES}/\sigma_{\ms RM} \equiv \sqrt{-2\ln\left\{ 1 - \left[  \tnm{erf}
		\left(\frac{1}{\sqrt{2}}\,\frac{|G|}{\sigma_{\ms RM}}\right)\right]^{\ms \!1/M}
		\right\}} \,.
\end{equation}
Examples illustrating use of these equations are presented in \S~\ref{sec:ex}.
Equation~(\ref{eqn:GtoORay}), and thus implicitly equation~(\ref{eqn:ORaytoG}),
is displayed for several values of $M$ in Fig.~\ref{fig:detsig}.

\section{Examples}\label{sec:ex}

In \S~\ref{sec:exL} and \S~\ref{sec:exLrm} we construct examples to
demonstrate use of the signal-free expectation value equations and significance
relationships derived for $L$ in \S~\ref{sec:standardDS} and $L_{\ms RM}$
in \ref{sec:rmstatsDS}, respectively. In \S~\ref{sec:exMUFC} we describe how
the significance relationships for $L_{\ms RM}$ may be used to assess
Faraday dispersion spectra comprising multiple unresolved Faraday components.

\subsection{Standard Linear Polarization}\label{sec:exL}

The expectation value for a measurement of $L$ for an emission-free
($L_{\ms 0}=0$) line of sight is given by equation~(\ref{eqn:RayM});
equivalently, equation~(\ref{eqn:RayM}) returns the average observed
intensity for a spatial pixel situated away from real sources in an
image of standard linear polarization, namely
$\sim1.25 \sigma_{\ms Q,U}$ (e.g. see behaviour of dashed curves in
Fig.~\ref{fig:pntsrc}).

Using equation~(\ref{eqn:LtoG}), we find that the detection of a linearly
polarized source with SNR $L_{\ms Q,U}/\sigma_{\ms Q,U}=4.0$ is equivalent
in significance to the detection of a $\pm3.6\sigma_{\ms Q,U}$ source under
Gaussian statistics.

Using equation~(\ref{eqn:GtoL}) or the $M=1$ curve from Fig.~\ref{fig:detsig},
we find that a detection threshold of $L/\sigma_{\ms Q,U}=5.4$ must be
imposed in order to ensure that polarization detections have an equivalent
Gaussian significance in excess of $\pm5.0\sigma_{\ms Q,U}$ (i.e. greater than
99.99994\% confidence).

\subsection{Faraday-Space Linear Polarization}\label{sec:exLrm}

The expectation value for a measurement of $L_{\ms RM}$ for an
emission-free line of sight is given by equation~(\ref{eqn:OrderRayM});
its value depends on $M$. Equivalently, equation~(\ref{eqn:OrderRayM})
returns the average observed intensity for a spatial pixel situated away
from real sources in an image of peak Faraday-space linear polarization,
which for an example image with $M=30$ is found to be
$\sim2.78 \sigma_{\ms RM}$ (e.g. see behaviour of solid curves in
Fig.~\ref{fig:pntsrc}).

We now demonstrate the statistical significance relationships for
$L_{\ms RM}$ with $M=30$ using the same examples from \S~\ref{sec:exL}.

Using equation~(\ref{eqn:ORaytoG}), we find the detection of
a source with $L_{\ms RM}/\sigma_{\ms RM}=4.0$ to be equivalent
in significance to the detection of a $\pm2.6\sigma_{\ms Q,U}$
source under Gaussian statistics.

Using equation~(\ref{eqn:GtoORay}), we find that a detection
threshold of $L_{\ms RM}/\sigma_{\ms RM}=6.0$ is required
to ensure equivalent Gaussian significance in excess of
$\pm5.0\sigma_{\ms RM}$.

\subsection{Multiple Unresolved Faraday Components}\label{sec:exMUFC}

In \S~\ref{sec:rmstats} we derived the PDF for $L_{\ms RM}$ by
assuming that $|\mathcal{F}^{cln}(\phi)|$ contains no more than a
single unresolved Faraday component. While derivations of PDFs for
polarized intensity measurements drawn from more complicated
Faraday dispersion spectra remain beyond the scope of this work,
we note that the single Faraday component assumption was not
formally required to derive the significance relationships
presented in \S~\ref{sec:rmstatsDS}. Indeed, these relationships
are suitable for assessing the Gaussian equivalent significance
for any number of the available $M$ statistically independent
measurements in $|\mathcal{F}^{cln}(\phi)|$. This is because the
relationships in effect benchmark the significance of any
observed sample against the maximum theoretical noise sample
expected within $|\mathcal{F}^{cln}(\phi)|$. Therefore,
equations~(\ref{eqn:ORaytoG}) and (\ref{eqn:GtoORay}), as
demonstrated in \S~\ref{sec:exLrm}, may be used to
evaluate the Gaussian equivalent significance for each
candidate Faraday component in $|\mathcal{F}^{cln}(\phi)|$;
this practice will help to identify noise-induced
components in complex Faraday dispersion spectra
\citep[e.g. see data obtained by][]{2011ApJ...728...57L}.

\section{Source Profiles In Polarization Images}\label{sec:psp}

Two-dimensional images of linearly polarized intensity for $L$
or $L_{\ms RM}$ may be formed by calculating equation~(\ref{eqn:L})
or (\ref{eqn:LrmDEF}), respectively, for each independent spatial
pixel. In this section we illustrate cross-sectional profiles
for astronomical sources as observed in images of linear polarization, to
both demonstrate use of the equations derived earlier, and to briefly
outline challenges that need to be met for robust image-plane source extraction.
For demonstration we focus on the observation of sources with Gaussian morphologies;
such sources are typically encountered in radio astronomy because
of the well-approximated Gaussian nature of telescope point
spread functions.

\begin{figure}
 \includegraphics[trim = 0mm 0mm 0mm 0mm, clip, angle=-90, width=84mm]{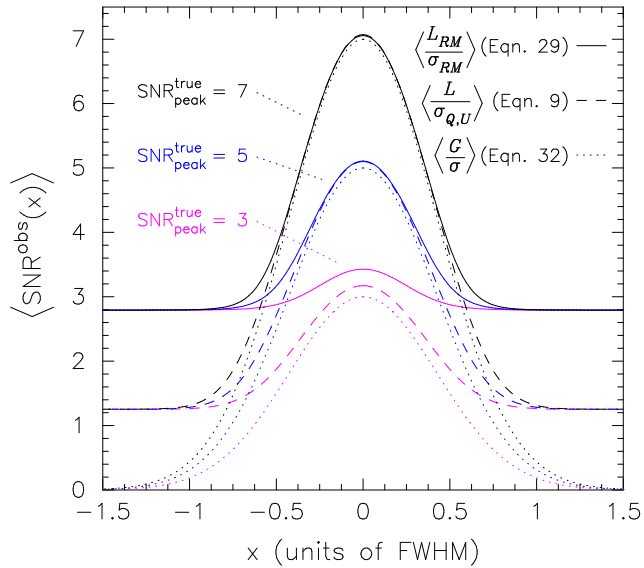}
 \caption{
	Mean spatial profiles observed for Gaussian sources embedded
	within images exhibiting Gaussian (dotted curves), $L$ (dashed
	curves), and $L_{\ms RM}$ (for $M=30$; solid curves) pixel intensity
	statistics, displayed for several input true peak SNRs. See
	\S~\ref{sec:psp} for details.}
 \label{fig:pntsrc}
\end{figure}

In Fig.~\ref{fig:pntsrc} we trace mean observed spatial profiles
through Gaussian sources, each with full width at half-maximum (FWHM)
standardised to unity, that have been embedded in images
for which the intensities of individual spatial pixels exhibit
the statistics of either Gaussian noise, the distribution
for $L$ from equation~(\ref{eqn:Rice}), or the distribution for
$L_{\ms RM}$ from equation~(\ref{eqn:Lrm}) with $M=30$;
we assume infinitesimal pixel dimensions so as to ignore pixel
discretisation effects.
The curves displayed in Fig.~\ref{fig:pntsrc} were obtained analytically
using the following approach. First, we constructed true underlying
SNR profiles for our Gaussian sources as spatial functions of fractional
FWHM, $x$, using
\begin{equation}\label{eqn:Gsrc}
	\tnm{SNR}^{\rm{true}}(x) = \tnm{SNR}^{\rm{true}}_{\rm{peak}}
	\exp\!\left[ - 4\ln\left(2\right)
	\frac{x^2}{\rm{FWHM}^2} \right] \;.
\end{equation}
This equation can be used to represent underlying cross-sectional
profiles for Gaussian sources in linear polarization, i.e. $L_{\ms 0}(x)$,
noting that Gaussian profiles in images of Stokes $Q$ and $U$ remain Gaussian through
equation~(\ref{eqn:L0}). To obtain observed spatial profiles for the
Gaussian noise (denoted by $G$), $L$, and $L_{\ms RM}$ images, we then
computed expectation values as a function of $x$ for the input signal
defined by equation~(\ref{eqn:Gsrc}). For the Gaussian noise profiles,
expectation values for observed SNRs equal their true underlying
SNRs (i.e. expectation values in Gaussian noise are signal-independent; i.e.
$\left<\tnm{SNR}^{\rm{obs}}(x)\right>=\tnm{SNR}^{\rm{true}}(x)\;,\;\forall x$).
The expectation values for $L(x)$ and $L_{\ms RM}(x)$ were computed
using equations~(\ref{eqn:RiceM}) and (\ref{eqn:LrmM}), respectively,
with $L_{\ms 0}(x)$ given by equation~(\ref{eqn:Gsrc}).
In signal-free regions (i.e. left and right of the sources),
$\left<L(x)\right>$ and $\left<L_{\ms RM}(x)\right>$ limit to
equations~(\ref{eqn:RayM}) and (\ref{eqn:OrderRayM}), respectively.
As $\tnm{SNR}^{\rm{true}}_{\rm{peak}}$ increases, the curves for
$\left<L(x)\right>$ and $\left<L_{\ms RM}(x)\right>$ limit to the
Gaussian case, the latter more slowly (cf. Fig.~\ref{fig:statsALL}).

Least squares 2D elliptical Gaussian fitting routines
\citep[e.g. the task {\tt IMFIT} from the {\tt MIRIAD} package;][]{1995ASPC...77..433S}
are typically used to extract sources from images exhibiting Gaussian noise.
The polarization profiles in Fig.~\ref{fig:pntsrc} suggest that
Gaussian fitting routines may not be appropriate for source extraction in
images of linear polarization, unless low SNR wings are excluded from the
fitting process (e.g. by imposing a SNR cut-off threshold for fitting).
Additionally, the non-Gaussian distribution of pixel intensities about the
mean profiles illustrated in Fig.~\ref{fig:pntsrc} will likely cause a
systematic positive bias in extracted flux densities, particularly for
low SNR sources. To address these challenges, detailed inspections regarding
the accuracy of source extraction methods in images of linear polarization
are required. Such analysis is beyond the scope of this work; see
\citet{halesB} for an analysis of source extraction in linear polarization.

\section{Non-Gaussian Noise in $Q/U$?}\label{sec:ngQU}

\citet{2011arXiv1106.5362G} recently suggested that aperture-synthesis
imaging and calibration artefacts may introduce strong non-Gaussianities
into the noise distribution for images of Stokes $Q$ and $U$,
which will in turn affect the false detection rate of sources in
linear polarization. To model these non-Gaussianities in Stokes $Q$
and $U$, they suggested use of a compound distribution comprising a
Gaussian distribution plus an exponential distribution; this is known
from the psychological literature as the Ex-Gaussian distribution
\citep[][]{hohle,burbeck}.

While it is likely that imaging artefacts will be present in images
of Stokes $Q$ and $U$, their influence should be largely accounted
for in {\it local} estimates of rms noise \citep[see e.g.][]{halesB}.
This process will ensure that the distribution of pixel SNRs is well
described by a Gaussian, in turn ensuring that local detection
thresholds can be computed accurately using the equations presented
in this paper.

In an effort to explain the seemingly non-Gaussian distribution exhibited
in the lower panel of Fig.~7 from \citet{2011arXiv1106.5362G}, in which
Stokes $Q$ data from the NVSS \citep{1998AJ....115.1693C} were presented,
we focus on two effects. We note that \citet{2011arXiv1106.5362G}
did not evaluate false detection rates using the NVSS data itself, but rather
a simulated sky survey described as having characteristics similar to the
NVSS. Here we examine whether the real NVSS data can be used to
justify claims of strong non-Gaussian noise.

\begin{figure}
 \includegraphics[clip, angle=-90, width=83mm]{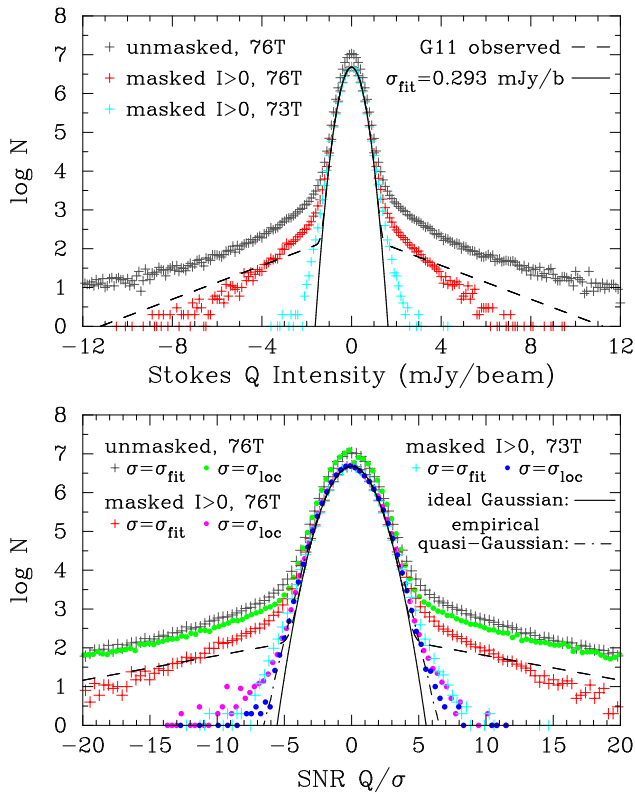}
 \caption{
	{\it Upper panel}: Distribution of Stokes $Q$ pixel intensities
	using all pixels (gray points) or only those with corresponding
	Stokes $I$ pixel intensities $\le0$~mJy~beam$^{-1}$ (red points)
	for our full sample of 76 NVSS tiles (76T). The cyan points represent
	the masked distribution for our 73 tile sample (73T), following removal
	of 3 tiles containing pointings with easily discerned imaging
	errors. The solid curve is a fitted Gaussian to the 73T points, with
	$\sigma_\tnm{\tiny fit}=0.293$~mJy~beam$^{-1}$. The dashed curve is from
	the lower panel of Fig.~7 from \citet{2011arXiv1106.5362G} for their
	observed NVSS data. {\it Lower panel}: Distribution of
	pixel SNRs corresponding to the points in the upper panel, obtained
	by assuming constant $\sigma=\sigma_\tnm{\tiny fit}$ or using
	local rms noise estimates $\sigma=\sigma_\tnm{\tiny loc}$. The
	solid curve is a Gaussian with unit variance. The dot-dashed curve,
	obtained empirically, predicts the distribution of pixel SNRs that
	will be observed when rms noise values exhibit $10\%$ error. The
	dashed curve represents the corresponding upper panel curve,
	normalised by $\sigma_\tnm{\tiny fit}$. Note that x-axes are
	not matched between upper and lower panels.
	}
 \label{fig:NVSS}
\end{figure}

The first consideration is the presence of real sources, which may be positive
or negative in Stokes $Q$ (or $U$). Such sources need to be masked prior to
investigation of the noise distribution. \citet{2011arXiv1106.5362G} masked
real total intensity sources out of the Stokes $Q$ NVSS data investigated. We
attempted to recover a non-Gaussian distribution using NVSS data by
investigating a sample of $4^\circ\times4^\circ$ tiles
selected to have central positions located along a line of constant
declination with J2000 $\delta=+28^\circ$; for simplicity we did not analyse
the full 2326 tiles comprising the NVSS. We selected 75 of the 90 tiles in this
declination range, avoiding 15 tiles containing missing pointings in
Stokes $Q$. We supplemented this sample with an additional tile, {\tt C1232P12},
chosen arbitrarily to ensure that at least one tile containing pointings with
significant amplitude calibration errors was included in our analysis. Thus
our raw data sample consisted of 76 tiles. We also investigated a subset of
73 of these tiles following the removal of tile {\tt C1232P12}, as well
as two other tiles, {\tt C0432P28} and {\tt C0448P28}, which were found to
contain pointings with minor yet distinct calibration errors.
The declination range above was selected to be
representative of the NVSS, comprising tiles positioned from the North
Galactic Pole down to and below the Galactic plane. Tiles near the
Galactic plane are likely contain large-scale emission unresolved by the NVSS,
which may in turn plausibly introduce non-Gaussianities into the data due to
difficulties encountered during deconvolution \citep[e.g.][]{1999ASPC..180..151C}.
To obtain Stokes $Q$ images as free from true sources as possible, we
conservatively masked all pixels that had corresponding Stokes $I$
intensities $>0$~mJy~beam$^{-1}$. The rms noise in the NVSS is
$\sigma_{\ms I}\approx0.45$~mJy~beam$^{-1}$ in Stokes $I$ and
$\sigma_{\ms Q}\approx0.29$~mJy~beam$^{-1}$ in Stokes $Q$. Therefore,
we note that if masking were only applied to pixels corresponding to
catalogued NVSS sources, namely pixels with
$I$~{\footnotesize $\gtrsim$}~$4.5\sigma_{\ms I}$, then real Stokes $Q$
emission from sources with $\sim20\%$ fractional polarization could
remain unmasked with significance up to $Q\approx1.4\sigma_{\ms Q}$,
biasing efforts to uncover the underlying noise distribution. We then
compared histograms of Stokes $Q$ pixels intensities for the unmasked and
masked data, and for the distribution observed by \citet{2011arXiv1106.5362G},
as shown in the upper panel of Fig.~\ref{fig:NVSS}.
We found that the unmasked 76 tile data displayed non-Gaussian wings above the NVSS levels
reported by \citet{2011arXiv1106.5362G}, while the masked 76 tile data displayed wings
that fell off more rapidly with intensity than their levels.
The 73 tile data was found to exhibit significantly attenuated wings compared with the
76 tile data. The distribution described by \citet{2011arXiv1106.5362G} is inconsistent
with the 73 tile data. The difference between the unmasked and masked 76 tile data in
Fig.~\ref{fig:NVSS} is due to real sources, while the difference between the masked 76
and 73 tile data is due to the inclusion of tiles with significant imaging errors.
Therefore, the discrepancy between the NVSS distribution observed by
\citet{2011arXiv1106.5362G} and the masked 73 tile NVSS distribution examined in
this work is likely to be due to residual unmasked sources, the inclusion of corrupted
tiles, a combination of both, or some other processing error.

To explain why the masked 73 tile data exhibit small non-Gaussian wings, we
now consider a second effect that may also lead to spurious claims of
non-Gaussian noise. It is common for rms noise to vary spatially throughout
an image due to imaging artefacts about strong sources, or intrinsic
observational features such as primary beam sensitivity. If a histogram
of pixel intensities is used as a proxy to examine the noise distribution
within such an image, rather than a histogram of pixel SNRs (which
require local rms noise estimates), then the inferred noise distribution
will appear to follow a Gaussian distribution with exponential wings.
Similarly, if the distribution of pixel SNRs is examined for these images
whilst assuming that rms noise is spatially uniform, then the inferred noise
distribution will again appear to follow the Ex-Gaussian distribution. In
both these examples, no intrinsic non-Gaussianities need exist. To
demonstrate, consider the following illustrative scenario
in which an image is arbitrarily divided into two spatial regions, each
represented by Gaussian statistics but with a different standard deviation.
If 99.9\% of the image has standard deviation 1 (in arbitrary
units) and 0.1\% has standard deviation 5 (this is a crude representation of
the fraction of 76 Stokes $Q$ tiles exhibiting calibration errors),
then the observed distribution of all pixel intensities
will follow an Ex-Gaussian distribution. Despite this suggested appearance
of non-Gaussianity, the distribution of pixel SNRs, obtained using
local rms noise estimates, will exhibit purely Gaussian
characteristics. Indeed, this simplified example further demonstrates
that if an observed distribution of pixel SNRs is not Gaussian
(following masking of real sources), then the implemented local rms
noise estimation procedure may not be performing suitably.

To construct distributions of pixel SNRs for the unmasked
and masked NVSS data presented above, we used the rms estimation algorithm implemented
within the {\tt SExtractor} package \citep{1996A&AS..117..393B,2005astro.ph.12139H}
to generate a background rms noise map for each tile. We set the local mesh
size to $24\times24$~pixel$^2$, an area equivalent to $N_{\ms b}=50$ independent
resolution elements \citep[see][]{halesB}. Our estimates of local rms noise
therefore have uncertainty
$\{[1+0.75/(N_{\ms b}-1)]^{\ms 2}[1-N_{\ms b}^{\ms -1}]-1\}^{\ms 0.5}\approx10\%$
\citep[using an approximation to the uncertainty of the standard error
estimator, suitable for $N_{\ms b}>10$; see p.~63 of][]{johnson}, or greater if
many resolution elements in a given mesh contain true sources. We then computed
pixel SNRs using the local rms noise values. For comparison, we also computed
pixel SNRs by assuming a spatially uniform rms noise value for all tiles. This
value was obtained from a fit to the masked 73 tile pixel intensity data,
as indicated by the solid curve in the upper panel of Fig.~\ref{fig:NVSS}.
The resulting SNR histograms for the unmasked and masked NVSS data are presented
in the lower panel of Fig.~\ref{fig:NVSS}. We found that the distributions
constructed using the uniform noise level exhibited stronger non-Gaussian
wings than those constructed using local noise estimates (the former
are equivalent to the pixel intensity distributions presented in the upper
panel of Fig.~\ref{fig:NVSS}), indicating the presence of spatial variations
in image sensitivity. This is most clearly demonstrated by the masked 76 tile
data; the failure of this data to exhibit a pure Gaussian distribution when
using local noise estimates may be predominantly attributed to rapid changes
in image sensitivity near corrupted pointings, where the accuracy of the rms
noise estimation algorithm employed by {\tt SExtractor} is diminished.
The SNR distribution for the masked 73 tile data with local rms noise
estimates (blue dots) was found to closely follow a Gaussian distribution,
modulo two apparently non-Gaussian features. First, neglecting bins with
$|\tnm{SNR}|$~{\footnotesize $\gtrsim$}~$6$, the distribution was found to
broaden with SNR against that of a true Gaussian. This broadening is
predominantly due to variance in the rms noise estimates used to calculate SNRs,
rather than any intrinsic features of the pixel intensity data. To demonstrate,
we simulated a distribution of SNRs by drawing samples from a Gaussian with
unit variance and dividing each sample by a noise term that was itself drawn
from a Gaussian with unit mean and standard deviation $10\%$. The resulting
distribution is displayed in the lower panel of Fig.~\ref{fig:NVSS},
providing a close fit to the observed data. Second, the distribution was
found to exhibit 18 pixels with $|\tnm{SNR}|>7$. We examined
the NVSS image data to determine the origin of these discrepant pixels.
We found that 5 of these pixels were situated one pixel beyond the masking
boundary of a strong total intensity source, where the total intensity emission
was observed to drop suddenly to become negative due an adjacent
noise trough or possibly a cleaning artefact. These Stokes $Q$ pixels were
thus associated with unmasked real emission. The remaining 13
pixels were situated within pointings exhibiting image striping, consistent
with calibration errors; additionally, each of these pointings were situated
close the Galactic plane, with Galactic longitude $\sim80^\circ$ and
latitude $\sim-15^\circ$. Given these identifications, we conclude that
there is no significant evidence for non-Gaussian noise in the Stokes $Q$ NVSS
data once pointings exhibiting easily discerned calibration errors have been
removed. While we cannot rule out the presence of non-Gaussian noise in all
surveys, the lack of evidence for such noise in the NVSS sample analysed
here suggests that surveys with more sophisticated data reduction are unlikely
to be affected, and almost certainly not at the strong levels suggested by
\citet{2011arXiv1106.5362G}.

\section{Conclusions and Future Work}\label{sec:conc}

We have derived customised statistics to describe Faraday-space
measurements of linearly polarized intensity obtained using
RM synthesis. The equations presented enable objective
determination of the significance of polarization detections,
which will be useful for upcoming surveys of radio polarization.

We found that when the observation-specific
parameter $M$ was increased, larger detection thresholds were
required for $L_{\ms RM}$ to ensure that noise features were not
mistaken for real polarized emission (e.g. see middle panel of
Fig.~\ref{fig:statsALL}). We found this effect to
be exponential; $M$ needed to be increased by at least an order of
magnitude (e.g. from 10 to 100, or 1,000 to 10,000) before the new
detection threshold required to satisfy an original level of
statistical significance needed to be significantly raised.
Therefore, we conclude that it is of limited practical
use to tailor observational setups to minimise $M$, unless $M$ can be
reduced by at least an order of magnitude. A suitable strategy for
large-$M$ observations, such as those afforded by the $M>10^4$ capabilities
of facilities such as the Australia Telescope Compact Array and
Expanded Very Large Array, may be to first perform RM-synthesis
using reduced spectral resolution (i.e. by averaging spectral
channels to reduce $T$, and in turn $M$) in order to identify faint
polarized emission over a reduced $\pm\phi_{\ms max}$ range.

We also discussed source extraction in polarization images and the
importance of obtaining spatially-dependent rms noise estimates.

We have not discussed the derivation of confidence intervals for
polarization measurements, the setting of upper limits, or
polarization bias. While detailed inspection of these issues
remains beyond the scope of this work, we close
by briefly highlighting how our results may be used in the future to
address each of them, as follows.

Using the equations developed in this work, credible intervals
(the Bayesian equivalent of frequentist confidence intervals)
may be constructed using the technique presented by
\citet[][]{2006PASP..118.1340V}. Similarly, confidence bounds
(which depend on observed intensities) and upper limits
(which do not depend on observed intensities, but rather on
the detection process and Type II error minimisation) may be
evaluated using the techniques presented by
\citet{2010ApJ...719..900K}; a demonstration using
observational data will be presented by Hales et al. (in preparation).

Finally, we note that the different statistics exhibited by
$L$ and $L_{\ms RM}$ prevent the application of polarization
debiasing schemes designed for the former \citep[e.g.][]{1985A&A...142..100S,vla161}
from being applied to the latter. However, exceptions
may be suitable in the limit to polarized sources strong enough to display
a similar PDF in both $L_{\ms RM}$ and $L$ (e.g. compare curves in the lower and
upper panels of Fig~\ref{fig:statsALL}). In general, a polarization bias correction
scheme designed for $L_{\ms RM}$ would need to be parameterised
by a term such as $M$ from equation~(\ref{eqn:Mdef}), so as to
take into account the experiment-specific terms $\phi_{\ms max}$
and $\psi$. We note that the fixed, un-parameterised debiasing
schemes presented by \citet{2011arXiv1106.5362G}
and \citet{2012ApJ...750..139M} are therefore limited in applicable
experimental scope. Future investigations are clearly required to
resolved these issues. Instead of attempting to correct observed
flux densities for polarization bias alone, an alternate approach may be
to combine this correction with one for \citet{1913MNRAS..73..359E} bias,
which like polarization bias is always present. A demonstration
of this combined technique to remove both polarization and
Eddington bias will be presented by Hales et al. (in preparation).

\section*{Acknowledgments}

We thank Tim Cornwell and Samuel M\"{u}ller for helpful discussions.
We thank the anonymous referee for comments that led to the improvement of this paper.
C.~A.~H. acknowledges the support of an Australian Postgraduate Award
and a CSIRO OCE Scholarship. B.~M.~G. acknowledges the support of an
Australian Laureate Fellowship from the Australian Research Council
through grant FL100100114.

\appendix

\section{Unequal Noise in $Q/U$}\label{sec:appA}

Derivation of the PDF for $L$ with $\sigma_{\ms Q}\neq\sigma_{\ms U}$ requires
marginalising the Euclidean norm of a bivariate normal distribution over position angle,
resulting in a complicated analytic expression that is difficult to utilise
and is more easily obtained and analysed numerically. Given
that $\sigma_{\ms Q}\neq\sigma_{\ms U}$ may be encountered in observational
data, as highlighted in the examples below, in this Appendix
we investigate how $\sigma_{\ms Q,U}$ may be defined such that the
analytic equations presented in this work for $L$ and $L_{\ms RM}$ may
remain {\it approximately} valid. While it is not possible
to fully model a bivariate normal distribution with a single noise term
$\sigma_{\ms Q,U}$ when $\sigma_{\ms Q}\neq\sigma_{\ms U}$, we note that the only
region of PDF parameter space that needs to be accurately modelled is that of
the noise outlier population (i.e. {\footnotesize $\gtrsim$}$5\sigma$); in practice,
detection thresholds may be suitably defined using this population, such that
accurate modelling of the remaining parameter space is not required. We
therefore focus here on obtaining a rudimentary definition for $\sigma_{\ms Q,U}$
that, under certain conditions, may facilitate use of the analytic relationships
presented in this work.

Two examples of data exhibiting $\sigma_{\ms Q}\neq\sigma_{\ms U}$ are as follows.
First, consider images of Stokes $Q$ and $U$ in which a polarized source is present
with signal $Q_{\ms 0} \neq U_{\ms 0}$, as will be the case in general. Following
deconvolution, it is possible for residual sidelobes and other artefacts about
strong sources to affect one image more than the other, causing some
lines-of-sight to exhibit $\sigma_{\ms Q}\neq\sigma_{\ms U}$. As a second example,
spatial variations in root-mean-square (rms) noise may be present and independently-positioned
throughout images of Stokes $Q$ and $U$, where beam-sized noise elements are
superposed on larger-scale undulations. For example, undulations in rms
noise may be produced in aperture synthesis images by large scale emission that
is unrecoverable by deconvolution algorithms such as {\tt CLEAN}
\citep{1999ASPC..180..151C}, or in general radio imaging through insufficient
flagging of data affected by radio frequency interference. Even surveys designed
to exhibit spatially uniform rms noise, such as the NVSS \citep{1998AJ....115.1693C}, 
exhibit undulations in rms noise due to a combination of the effects described
above, thus enabling some lines of sight to exhibit $\sigma_{\ms Q}\neq\sigma_{\ms U}$.
Though the issues above may be mitigated by telescope design, observing strategy,
and data processing, the potential remains for lines-of-sight to exhibit 
$\sigma_{\ms Q}\neq\sigma_{\ms U}$. Additionally, and in a
trivial sense, uncertainties in the estimator used to evaluate $\hat{\sigma}_{\ms Q}$ and
$\hat{\sigma}_{\ms U}$ (using hat notation here to indicate standard errors
rather than true underlying standard deviations) will result in
$\hat{\sigma}_{\ms Q}\neq\sigma_{\ms Q}$ and $\hat{\sigma}_{\ms U}\neq\sigma_{\ms U}$,
such that $\hat{\sigma}_{\ms Q}\neq\hat{\sigma}_{\ms U}$ may result in situations where
$\sigma_{\ms Q}=\sigma_{\ms U}$. For example, if a mesh containing $N_{\ms b}<100$
independent resolution elements is used to estimate the local standard error in an
image \citep[e.g. as demonstrated in Stokes $I$ by][]{2005AJ....130.1373H}, then the
uncertainty in this estimator will be $>7\%$ (using the formula
referenced in-text in \S~\ref{sec:ngQU}). Given this and
the examples above, how should $\sigma_{\ms Q,U}$ be defined? Ideally, when
$\sigma_{\ms Q}\neq\sigma_{\ms U}$, the use of $\sigma_{\ms Q,U}$ should be avoided
altogether and all analysis should be conducted numerically to correctly utilise
the true PDF. However, this may be cumbersome for typical situations where
$\hat{\sigma}_{\ms Q}$ and $\hat{\sigma}_{\ms U}$ are within a factor of, say,
$\sim10\%$.

We begin by focusing on $L$ and noting that in situations where
$\sigma_{\ms Q}=\sigma_{\ms U}$, a suitable definition for $\sigma_{\ms Q,U}$ may
be given by $0.5(\hat{\sigma}_{\ms Q}+\hat{\sigma}_{\ms U})$; this solution is
more precise (i.e. exhibits less dispersion about the true standard deviation)
than assigning $\sigma_{\ms Q,U}=\hat{\sigma}_{\ms Q}$ (or
$\sigma_{\ms Q,U}=\hat{\sigma}_{\ms U}$), because the probability of
$\hat{\sigma}_{\ms Q}$ overestimating the true standard deviation is greater
than the probability of both $\hat{\sigma}_{\ms Q}$ and $\hat{\sigma}_{\ms U}$
overestimating it. Similar performance may be obtained by defining
$\sigma_{\ms Q,U}$ following first-order error propagation \citep[e.g.][]{clif}
evaluated about the point $(Q_{\ms 0},U_{\ms 0})$,
\begin{equation}\label{eqn:sigDEF}
	\sigma_{\ms Q,U}^{\ms 2} \approx
	\left[\frac{\partial L}{\partial Q}\left(Q_{\ms 0}\right)\right]^{\!\ms 2}\sigma_{\ms Q}^{\ms 2} +
	\left[\frac{\partial L}{\partial U}\left(U_{\ms 0}\right)\right]^{\!\ms 2}\sigma_{\ms U}^{\ms 2} \;.
\end{equation}
We note that neglection of higher-order terms in the equation above
is formally incorrect (such terms are important when noise dominates signal); however,
our rudimentary interest here regards assessment of the general first-order form of
equation~(\ref{eqn:sigDEF}), rather than its detailed quantitative properties.
Equation~(\ref{eqn:sigDEF}) indicates that a suitable first-order form for
$\sigma_{\ms Q,U}$ may be defined by
\begin{equation}\label{eqn:Lsig}
	\sigma_{\ms Q,U}^{\ms 2} \equiv
		A_{\ms Q}\hat{\sigma}_{\ms Q}^{\ms 2}+A_{\ms U}\hat{\sigma}_{\ms U}^{\ms 2},
\end{equation}
with positive factors $A_{\ms Q}$ and $A_{\ms U}$ satisfying $A_{\ms Q}+A_{\ms U}=1$,
and where these factors may be assumed to be constants (i.e.
independent of $Q_{\ms 0}$ and $U_{\ms 0}$) to ensure that $\sigma_{\ms Q,U}$ remains
signal-independent (note comments at the end of \S~\ref{sec:standard}).
To investigate whether the use of equation~(\ref{eqn:Lsig}), or the average standard
error definition further above, could enable outliers within empirically-obtained
$\sigma_{\ms Q}\neq\sigma_{\ms U}$ distributions to be suitably characterised by
the analytic distributions for $L$ and $L_{\ms RM}$, we performed the following
simulations.

We populated discrete distributions of $L/\sigma_{\ms Q,U}$ (dimensionless) for
$1 \le \sigma_{\ms Q}/\sigma_{\ms U} \le 1.5$ and examined how closely outliers beyond
$\sim5\sigma$ were fit by equation~(\ref{eqn:Ray}). Three definitions for
$\sigma_{\ms Q,U}$ were tested: $0.5(\hat{\sigma}_{\ms Q}+\hat{\sigma}_{\ms U})$,
equation~(\ref{eqn:Lsig}) with $A_{\ms Q}=A_{\ms U}=0.5$, and a case where $A_{\ms Q}$ and
$A_{\ms U}$ were varied in search of optimal values. We set up our simulations in two
ways. In the first setup, we assumed that the variance in standard error estimates was
zero, namely that $\hat{\sigma}_{\ms Q}=\sigma_{\ms Q}$ and $\hat{\sigma}_{\ms U}=\sigma_{\ms U}$.
We found that selecting $A_{\ms Q}=1$ or $A_{\ms U}=1$ over- or underestimated the noise
required to correctly model true outlier populations, respectively. Selecting these values in turn
under- or overestimated the true statistical significance of outliers, respectively. We therefore
note that if we conservatively defined
$\sigma_{\ms Q,U} \equiv \max{\left(\sigma_{\ms Q},\sigma_{\ms U}\right)}$,
then the statistical significance of outliers, as well as that of true polarized
sources, would never be overestimated (though they would certainly be underestimated).
To limit the degree to which the significance of polarization detections (both
signal and noise) could be underestimated, whilst still generally preventing them
from being overestimated, we found the following empirical values to be suitable
for use in equation~(\ref{eqn:Lsig}):
\begin{equation}\label{eqn:Afactors}
  A_{\ms Q} = \left\{
  \begin{array}{l l}
    0.8 & \quad \tnm{if $\sigma_{\ms Q} \ge \sigma_{\ms U}$}\\
    0.2 & \quad \tnm{if $\sigma_{\ms Q} < \sigma_{\ms U}$}\;,\\
  \end{array} \right.
\end{equation}
with $A_{\ms U}=1-A_{\ms Q}$. We found that use of these factors limited
systematic underestimation of the statistical significance of
$G^{\ms ES}\approx5\sigma$ outlier samples to {\footnotesize $\lesssim$}~2\%
of their true, empirically determined values;
this underestimation peaked at $\sigma_{\ms Q}/\sigma_{\ms U}\sim1.2$,
diminishing elsewhere within the $1 \le \sigma_{\ms Q}/\sigma_{\ms U} \le 1.5$ range
tested. For comparison, we found that the conservative definition
$\sigma_{\ms Q,U} \equiv \max{\left(\sigma_{\ms Q},\sigma_{\ms U}\right)}$
performed worse, typically underestimating true significance values systematically
by $\sim6\%$. In the second simulation setup, we investigated
the effects of uncertainties in $\hat{\sigma}_{\ms Q}$ and $\hat{\sigma}_{\ms U}$
by assuming what we envisaged to be $\sim$worst-case 10\%
errors in each. The factors from equation~(\ref{eqn:Afactors}) were again found
to be suitable in these simulations.

\begin{figure*}
 \includegraphics[clip, angle=-90, width=160mm]{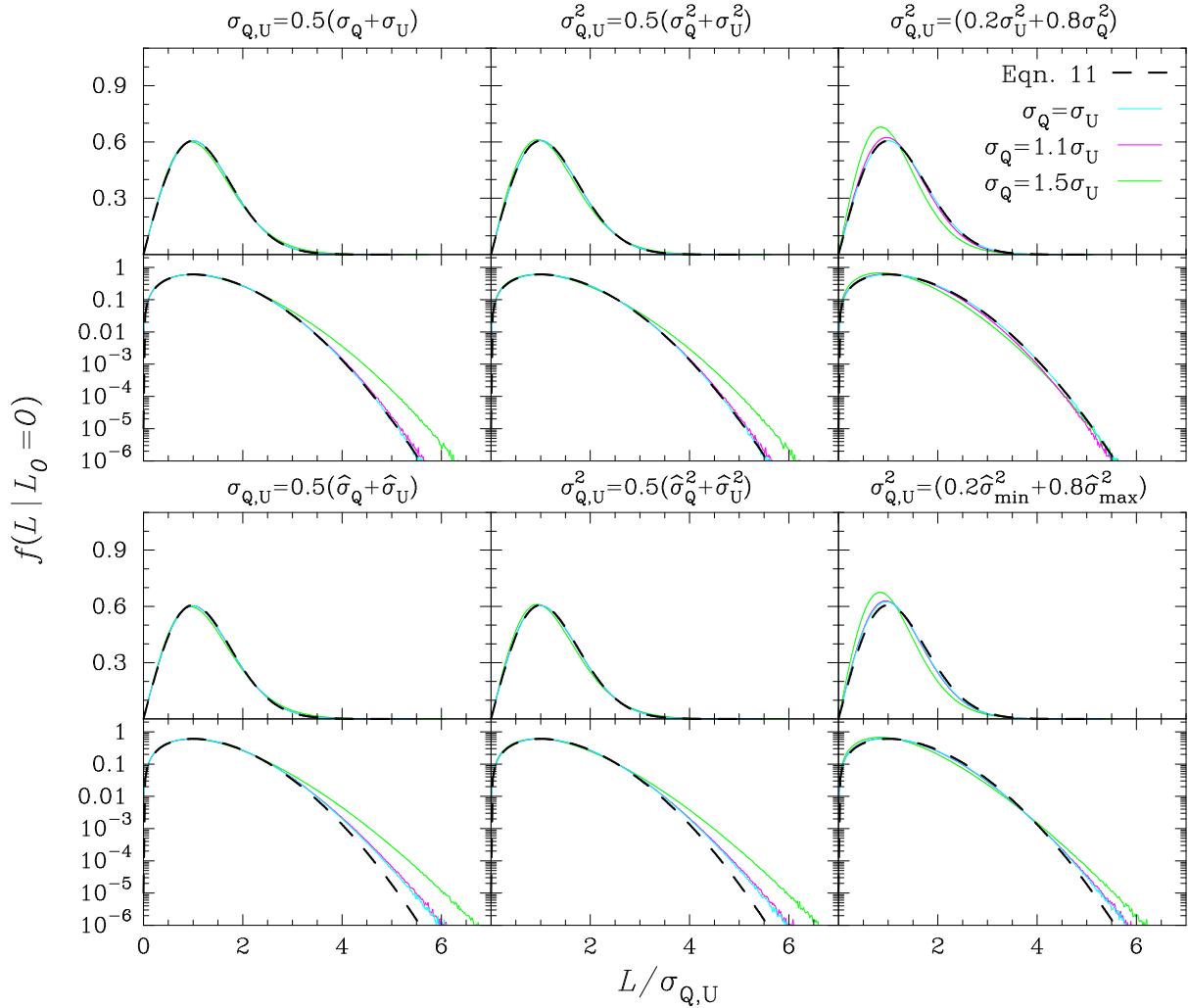}
 \caption{
	Empirical PDFs for $L$ obtained for three noise cases (coloured curves),
	in which each $L$ sample was normalised by $\sigma_{\ms Q,U}$ in accordance
	with the three definitions indicated above the columns. For comparison,
	the analytic PDF for $L$ given by equation~(\ref{eqn:Ray}) is presented
	identically in each panel; it is not fit to the data. Data from panels
	in the first and third rows are displayed with logarithmic scaling in the
	second and fourth rows, respectively. The upper set of panels are for the
	simulations with known standard errors, namely for
	$\hat{\sigma}_{\ms Q}=\sigma_{\ms Q}$ and
	$\hat{\sigma}_{\ms U}=\sigma_{\ms U}$. The lower set of panels are for
	the simulations with 10\% errors in $\hat{\sigma}_{\ms Q}$
	and $\hat{\sigma}_{\ms U}$. Small-number statistics begin to cause artificial
	broadening of the empirical distributions below densities of $\sim10^{-5}$.
}
 \label{fig:NEsigQUL}
\end{figure*}

The results of our $L$ simulations are summarised in Fig.~\ref{fig:NEsigQUL} for
the three $\sigma_{\ms Q,U}$ definitions tested, and for three noise cases:
$\sigma_{\ms Q}=\sigma_{\ms U}$, $\sigma_{\ms Q}=1.1\sigma_{\ms U}$, and
$\sigma_{\ms Q}=1.5\sigma_{\ms U}$.
We found that introducing variance into the standard error estimates affected the
empirical $L$ distributions in a manner similar to the influence of introducing
$\sigma_{\ms Q}\neq\sigma_{\ms U}$, shifting outlier populations away from the
analytic PDFs. We found that the $0.5(\hat{\sigma}_{\ms Q}+\hat{\sigma}_{\ms U})$
and $A_{\ms Q}=A_{\ms U}=0.5$ definitions for $\sigma_{\ms Q,U}$ resulted in
empirical distributions that were closely fit by equation~(\ref{eqn:Ray}) at
low-SNRs, but which resulted in increasingly poor fits for outlier populations
as either variance was introduced to the standard error estimates or
$\sigma_{\ms Q}\neq\sigma_{\ms U}$ was introduced. The $A_{\ms Q}=A_{\ms U}=0.5$
definition resulted in marginally improved fits for noise outliers compared
with the average standard error definition.
The third definition, using equation~(\ref{eqn:Lsig}) with factors from
equation~(\ref{eqn:Afactors}), resulted in empirical distributions most
accurately fit by equation~(\ref{eqn:Ray}) for outliers, though at the expense
of poor fitting at low-SNR. We found that this definition of $\sigma_{\ms Q,U}$
optimally accounted for unequal noise within the range tested, even when
variance in standard error estimates was introduced (though less
so towards the upper end of the range where $\sigma_{\ms Q}/\sigma_{\ms U}=1.5$).

We performed similar simulations for $L_{\ms RM}$ to investigate suitability
of the $\sigma_{\ms Q,U}$ definitions considered above in this different
statistical environment. We simplified the potential complexity
of these simulations by assuming the following illustrative $\sim$worst-case setup
based on the $M=30$, 24 spectral channel setup described in \S~\ref{sec:rmstats}.
Individual samples of $L_{\ms RM}$ were obtained empirically
by selecting the maximum of $M=30$ independent Rayleigh-distributed variates.
Each of these variates was assumed to represent a uniformly-weighted stack
of 24 spectral channels, namely with equation~(\ref{eqn:RMweights}) set to unity.
We investigated unequal noise by systematically assuming
$\sigma_{\ms Q,i}=\sigma_{\ms U,i}$, $\sigma_{\ms Q,i}=1.1\sigma_{\ms U,i}$, or
$\sigma_{\ms Q,i}=1.5\sigma_{\ms U,i}$ for each $i$'th channel, and investigated
the influence of uncertainties in estimates of standard error by introducing
10\% error to $\hat{\sigma}_{\ms Q,i}$ and
$\hat{\sigma}_{\ms U,i}$ for each channel (thus $10/\sqrt{24}$\%
in each of the $30\times2$ stacked $Q$ and $U$ values used to obtain each
sample of $L_{\ms RM}$).

\begin{figure*}
 \includegraphics[clip, angle=-90, width=160mm]{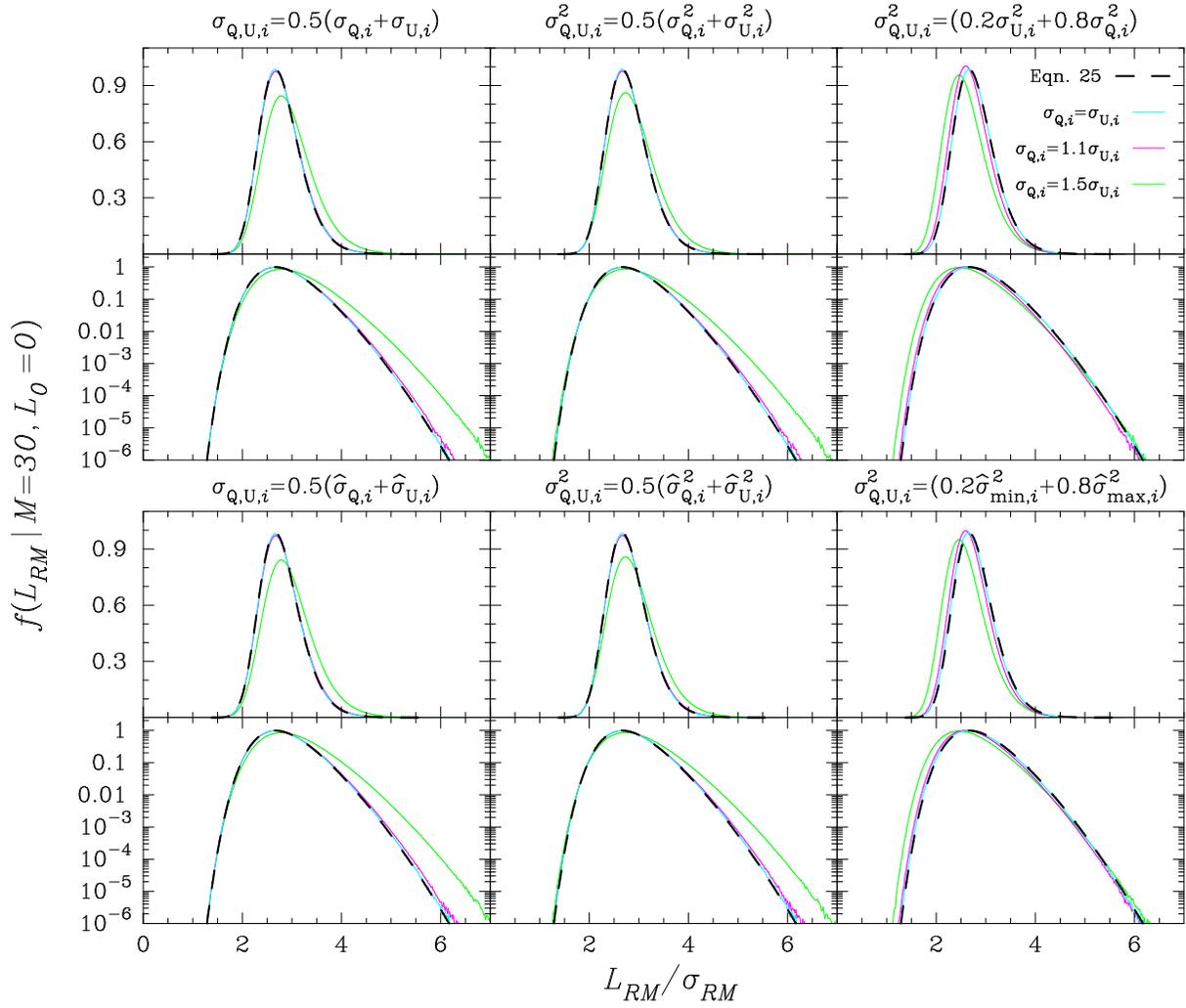}
 \caption{
	Empirical PDFs for $L_{\ms RM}$ compared with the analytic PDF
	given by equation~(\ref{eqn:OrderRay}); see text for details.
	The layout follows that described for Fig.~\ref{fig:NEsigQUL}.
}
 \label{fig:NEsigQULrm}
\end{figure*}

The results of our $L_{\ms RM}$ simulations are displayed in Fig.~\ref{fig:NEsigQULrm}.
We found similar behaviour of the different $\sigma_{\ms Q,U}$ definitions to
that exhibited in Fig.~\ref{fig:NEsigQUL}. However, the effect of introducing
unequal noise was found to be more prominent within outlier populations
in Fig.~\ref{fig:NEsigQULrm} due to their being selected from a maximum of
$M=30$ samples. This effect will be further pronounced for data with larger
$M$, though because it is unlikely that all spectral channels will exhibit the
systematic unequal noise assumed in our worst-case simulations, the outlier
populations in real data are unlikely to be as extreme as those presented here.
Additionally, if channel weighting is introduced according to
equation~(\ref{eqn:RMweights}), the the impacts of bad channel data will be
minimised and outlier populations further reduced. Regarding the introduction
of variance in $\hat{\sigma}_{\ms Q,i}$ and $\hat{\sigma}_{\ms U,i}$, we note
that its impact on outlier populations will be minimised for data with
a greater number of spectral channels, due to the $1/\sqrt{T}$ dependence
in stacked channel data. As with the simulations for $L$, we found that
by defining $\sigma_{\ms Q,U}$ using equation~(\ref{eqn:Lsig}) with the
factors from equation~(\ref{eqn:Afactors}), outlier populations of the various
empirical distributions investigated were aligned robustly with the analytic
distribution given by equation~(\ref{eqn:OrderRay}).

We conclude that the analytic equations presented in the main body of this
paper may be utilised for data exhibiting modest unequal noise of $\sim10\%$,
without incurring significant biases, by defining $\sigma_{\ms Q,U}$ according to
equation~(\ref{eqn:Lsig}) with factors from equation~(\ref{eqn:Afactors}).
While our simulations have indicated that this definition may be suitable for
data exhibiting more extreme unequal noise, we caution that we have not
considered the potential impacts of biases on polarization position angles;
future investigation regarding this issue is clearly required.

\bsp

\label{lastpage}


\begin{thebibliography}{99}

\bibitem[\protect\citeauthoryear{Anderson}{1971}]{anderson}
Anderson T.~W., 1971, The Statistical Analysis of Time Series.
John Wiley \& Sons, New York, NY

\bibitem[\protect\citeauthoryear{Bertin \& Arnouts}{1996}]{1996A&AS..117..393B}
Bertin E., Arnouts S., 1996, A\&AS, 117, 393

\bibitem[\protect\citeauthoryear{Brentjens \& de Bruyn}{2005}]{2005A&A...441.1217B}
Brentjens M.~A., de Bruyn A.~G., 2005, A\&A, 441, 1217 

\bibitem[\protect\citeauthoryear{Burbeck \& Luce}{1982}]{burbeck}
Burbeck S.~L., Luce R.~D., 1982, Perception and Psychophysics, 32, 117

\bibitem[\protect\citeauthoryear{Burn}{1966}]{1966MNRAS.133...67B}
Burn B.~J., 1966, MNRAS, 133, 67

\bibitem[\protect\citeauthoryear{Clifford}{1973}]{clif}
Clifford A.~A., 1973, Multivariate Error Analysis. Applied Science Publishers, London

\bibitem[\protect\citeauthoryear{Condon et al.}{1998}]{1998AJ....115.1693C}
Condon J.~J., Cotton W.~D., Greisen E.~W., Yin Q.~F., Perley R.~A., Taylor 
G.~B., Broderick J.~J., 1998, AJ, 115, 1693

\bibitem[\protect\citeauthoryear{Cornwell, Braun, \& Briggs}{1999}]{1999ASPC..180..151C}
Cornwell T., Braun R., Briggs D.~S., 1999, ASPC, 180, 151

\bibitem[\protect\citeauthoryear{David \& Nagaraja}{2003}]{david}
David H.~A., Nagaraja H.~N., 2003, Order Statistics (3rd ed.). John Wiley \& Sons,
Hoboken, NJ

\bibitem[\protect\citeauthoryear{Deboer et al.}{2009}]{2009IEEEP..97.1507D}
Deboer D.~R., et al., 2009, IEEE Proceedings, 97, 1507

\bibitem[\protect\citeauthoryear{Eddington}{1913}]{1913MNRAS..73..359E}
Eddington A.~S., 1913, MNRAS, 73, 359

\bibitem[\protect\citeauthoryear{Gaensler et al.}{2010}]{2010AAS...21547013G}
Gaensler B.~M., Landecker T.~L., Taylor A.~R., POSSUM Collaboration,
2010, BAAS, 42, 515

\bibitem[\protect\citeauthoryear{George et al.}{2011}]{2011arXiv1106.5362G}
George S.~J., Stil J.~M., Keller B.~W., 2011, PASA, in press (arXiv:1106.5362)

\bibitem[\protect\citeauthoryear{Hales et al.}{2012}]{halesB}
Hales C.~A., Murphy T., Curran J.~R., Middelberg E., Gaensler B.~M.,
Norris R.~P., 2012, MNRAS, in press

\bibitem[\protect\citeauthoryear{Heald et al.}{2009}]{2009A&A...503..409H}
Heald G., Braun R., Edmonds R., 2009, A\&A, 503, 409

\bibitem[\protect\citeauthoryear{Hohle}{1965}]{hohle}
Hohle R.~H., 1965, Journal of Experimental Psychology, 69, 382

\bibitem[\protect\citeauthoryear{Holwerda}{2005}]{2005astro.ph.12139H}
Holwerda B.~W., 2005, arXiv:astro-ph/0512139

\bibitem[\protect\citeauthoryear{Huynh et al.}{2005}]{2005AJ....130.1373H}
Huynh M.~T., Jackson C.~A., Norris R.~P., Prandoni I., 2005, AJ, 130, 1373

\bibitem[\protect\citeauthoryear{Johnson \& Kotz}{1970}]{johnson}
Johnson N.~L., Kotz S., 1970, Distributions in Statistics: Continuous Univariate
Distributions$-$1. Houghton Mifflin, NY

\bibitem[\protect\citeauthoryear{Johnston et al.}{2008}]{2008ExA....22..151J}
Johnston S., et al., 2008, Experimental Astronomy, 22, 151

\bibitem[\protect\citeauthoryear{Kashyap et al.}{2010}]{2010ApJ...719..900K}
Kashyap V.~L., van Dyk D.~A., Connors A., et al., 2010, ApJ, 719, 900

\bibitem[\protect\citeauthoryear{Law et al.}{2011}]{2011ApJ...728...57L}
Law C.~J., Gaensler B.~M., Bower G.~C., et al., 2011, ApJ, 728, 57

\bibitem[\protect\citeauthoryear{Leahy \& Fernini}{1989}]{vla161}
Leahy P., Fernini I., 1989, VLA Scientific Memorandum No. 161, NRAO

\bibitem[\protect\citeauthoryear{Macquart et al.}{2012}]{2012ApJ...750..139M}
Macquart J.-P., Ekers R.~D., Feain I., Johnston-Hollitt M., 2012, ApJ, 750, 139

\bibitem[\protect\citeauthoryear{Marcum}{1948}]{marcum}
Marcum J.~I., 1948, RAND Research Memo, RM-753, (Reprinted in 1960,
IRE Transactions on Information Theory, IT-6, 59)

\bibitem[\protect\citeauthoryear{Perley et al.}{2011}]{2011ApJ...739L...1P}
Perley R.~A., Chandler C.~J., Butler B.~J., Wrobel J.~M., 2011, ApJL, 739, L1

\bibitem[\protect\citeauthoryear{Rayleigh}{1880}]{rayleigh}
Rayleigh J.~W.~S., 1880, Philosophical Magazine, 5th Series, 10, 73

\bibitem[\protect\citeauthoryear{Rice}{1945}]{rice}
Rice S.~O., 1945, Bell System Technical Journal, 24, 46; Reprinted
by Wax N., 1954, Selected Papers on Noise and Stochastic Processes.
Dover Publications, NY, p. 133

\bibitem[\protect\citeauthoryear{Sault et al.}{1995}]{1995ASPC...77..433S}
Sault R.~J., Teuben P.~J., Wright M.~C.~H., 1995, Astronomical Data
Analysis Software and Systems IV, 77, 433

\bibitem[\protect\citeauthoryear{Simmons \& Stewart}{1985}]{1985A&A...142..100S}
Simmons J.~F.~L., Stewart B.~G., 1985, A\&A, 142, 100

\bibitem[\protect\citeauthoryear{Simon}{1998}]{simon}
Simon M.~K., 1998, IEEE Commun. Lett., 2, 39


\bibitem[\protect\citeauthoryear{Taylor \& Salter}{2010}]{2010ASPC..438..402T}
Taylor A.~R., Salter C.~J., 2010, in Kothes R., Landecker T. L., Willis
A. G., eds, ASP Conf. Ser. Vol. 438, The Dynamic Interstellar Medium: A
Celebration of the Canadian Galactic Plane Survey. Astron. Soc. Pac.,
San Francisco, p. 402

\bibitem[\protect\citeauthoryear{Vaillancourt}{2006}]{2006PASP..118.1340V}
Vaillancourt J.~E., 2006, PASP, 118, 1340 

\bibitem[\protect\citeauthoryear{Wilson et al.}{2011}]{2011MNRAS.416..832W} 
Wilson W.~E., et al., 2011, MNRAS, 416, 832

\end{thebibliography}
\end{document}